\documentclass[10pt,twocolumn,english,aps,manuscript,prb]{revtex4}
\usepackage[T1]{fontenc}
\usepackage[latin1]{inputenc}
\usepackage{babel}
\usepackage{graphics}

\makeatletter

\providecommand{\LyX}{L\kern-.1667em\lower.25em\hbox{Y}\kern-.125emX\@}

\usepackage[T1]{fontenc}
\usepackage[latin1]{inputenc}
\usepackage{babel}
\usepackage{amssymb}

\makeatletter
\date{}
\newcommand{\bra}[1]{{\langle #1 |}}
\newcommand{\ket}[1]{{| #1 \rangle}}
\newcommand{\up}{\hspace{-0.3 mm}\uparrow}
\newcommand{\down}{\hspace{-0.3 mm}\downarrow}

\newcommand{\then}{\Rightarrow}

\newcommand{\bbbone}{{\mathchoice {\rm 1\mskip-4mu l} {\rm 1\mskip-4mu l}{\rm 1\
mskip-4.5mu l} {\rm 1\mskip-5mu l}}}

\newcommand{\imag}{ {\rm I} \hspace{-1.0mm} {\rm Im} \:}

\makeatother
\begin{document}

\title{Coulomb scattering cross-section in a 2D electron gas and production
of entangled electrons }

\author{D. S. Saraga\( ^{1} \), B. L. Altshuler\( ^{2,3} \), Daniel Loss\( ^{1} \),
and R. M. Westervelt\( ^{4} \)}

\affiliation{{\footnotesize \( ^{1} \)Department of Physics and Astronomy, University
of Basel, Klingelbergstrasse 82, CH-4056 Basel, Switzerland }}

\affiliation{{\footnotesize \( ^{2} \)Physics Department, Princeton University,
Princeton, New Jersey 08544}}

\affiliation{{\footnotesize \( ^{3} \)NEC Research Institute, 4 Independence
Way, Princeton, New Jersey 08540}}

\affiliation{{\footnotesize \( ^{4} \)Division of Engineering and Applied Sciences,
Harvard University, Cambridge, Massachusetts 02138}}

\date{\( \today  \)}

\begin{abstract}
We calculate the Coulomb scattering amplitude for two electrons injected with opposite
momenta in an interacting 2DEG. We include the effect of the Fermi liquid background by
solving the 2D Bethe-Salpeter equation for the two-particle Green function vertex, in the
ladder and random phase approximations. This result is used to discuss the feasibility of
producing spin EPR pairs in a 2DEG by collecting electrons emerging from collisions at a
\( \pi /2 \) scattering angle, where only the entangled spin-singlets avoid the
destructive interference resulting from quantum indistinguishability. Furthermore, we
study the effective 2D electron-electron interaction due to the exchange of virtual
acoustic and optical phonons, and compare it to the Coulomb interaction.  Finally, we show
that the 2D Kohn-Luttinger pairing instability for the scattering electrons is negligible
in a GaAs 2DEG.
\end{abstract}
\maketitle

\section{Introduction}

Recent experiments \cite{Top00,Roy02} have allowed for the imaging of the coherent
electron flow in a two-dimensional electron gas (2DEG), demonstrating a roughly
directional injection through a quantum point contact (QPC) tuned to his lowest
transversal mode. We propose here to use such a setup to investigate Coulomb scattering in
2D, by measuring the scattering cross-section. This provides a natural motivation for
solving a long-lasting problem in Fermi-liquid theory: finding the scattering amplitude
for the Coulomb interaction in a 2D system, including the effect of the many-particle
background of the interacting Fermi sea. We derive \cite{Sar04} the scattering amplitude
\( f \) by solving the Bethe-Salpeter equation in the ladder approximation and for
electrons in the Cooper channel (opposite momenta) \cite{FW71,Mah00}.  This solution
provides a useful addition to Fermi-liquid theory applied to electron-electron
interaction. The development of Fermi-liquid theory, which goes back over many decades,
includes discussions of screening \cite{Ste67}, the lifetime of quasi-particles
\cite{Giu82,Zhe96}, the renormalization factor \( Z \) of the Green function
\cite{Bur00,Gal04}, the effective mass \cite{Tin75,Gal04}, and scattering
\cite{Lut66,Chu93,Suw04}.  An additional issue is how strongly the Coulomb scattering is
affected by the effective electron-electron interaction mediated by the exchange of
virtual phonons, which have been studied, for instance, in the context of Coulomb drag
(see e.g. Ref. \cite{Bon98}) and screening
\cite{Jal89}. Another extension concerns the effect of lower dimensionality
\cite{Chu93,Gal03} in Kohn-Luttinger superconductivity \cite{Lut66},
and the question of the strength of superconducting fluctuations (if
any), which could, in principle, spoil the Coulomb scattering. 

A second motivation for this work comes from the current efforts devoted
to solid-state implementations of quantum information protocols using
the spin of individual electrons as qubits \cite{Los98,BurLos,Van03}. In particular,
the experimental demonstration of entangled (EPR) pairs of spin-qubit
is still a present-day challenge, and has motivated a number of theoretical
proposals for \emph{entanglers}, i.e. devices creating mobile (spin-)
entangled electrons. These proposals relied on energy filtering via
quantum dots \cite{Los98,Rec01,Oli02,Sar03} and carbon nanotubes
\cite{Rec02}, and/or the use of superconductors \cite{Rec01,Les01,Rec02,Rec03}.
Other schemes are in closer relation with optics, and use beam-splitters
for spin \cite{Cos01}, orbital \cite{Sam03} or particle-hole entanglement
\cite{Bee03,Sam04}. 

We propose \cite{Sar04} here a simple idea for the creation of spin-entangled
pairs in a two-dimensional electron gas (2DEG), inspired by the well-known
interference effect found \emph{in vacuum} for the scattering of indistinguishable
particle \cite{Tay72,Jac01}. Using the fact that electron pairs
in the singlet (triplet) spin state behave like spinless bosons (fermions)
in spin-independent collisions, we propose to collect electrons emerging
from electron-electron collision with a scattering angle \( \theta =\pi /2 \).
In this situation, the destructive interference is complete for triplets;
hence the collected electrons must be in the entangled singlet state
\( \left| S\right\rangle =(\ket {\up \down }-\ket {\down \up })/\sqrt{2} \),
which is one the EPR states desirable for quantum information protocols.
The question now arises whether this two-particle exchange effect
survives in the presence of a sea of interacting electrons. Using
our solution for the scattering amplitude \( f \), we will show that
the entanglement created (or rather, post-selected) by the collision
should be observable in a realistic 2DEG. 

We emphasize that our study of electron-electron interaction is motivated 
by the prospect of seeing experimental control on the 
propagation of electrons in 2DEGs and on their
quasi-particle properties. The experiments of Ref.
\cite{Top00,Roy02} have used electron scattering off an SPM tip to
image both quantum interferences and ray-like propagation of electrons,
including diffraction. They have also shown control over the quasi-particle
lifetimes for hot electrons, in good agreement with Fermi-liquid theory.
The theoretical work presented here indicate that such experimental
studies can be extended to more general Fermi-liquid effects involving
two quasi-particles \cite{Anderson}.

\begin{figure*}[t]
{\centering \resizebox*{0.9\textwidth}{!}{\includegraphics{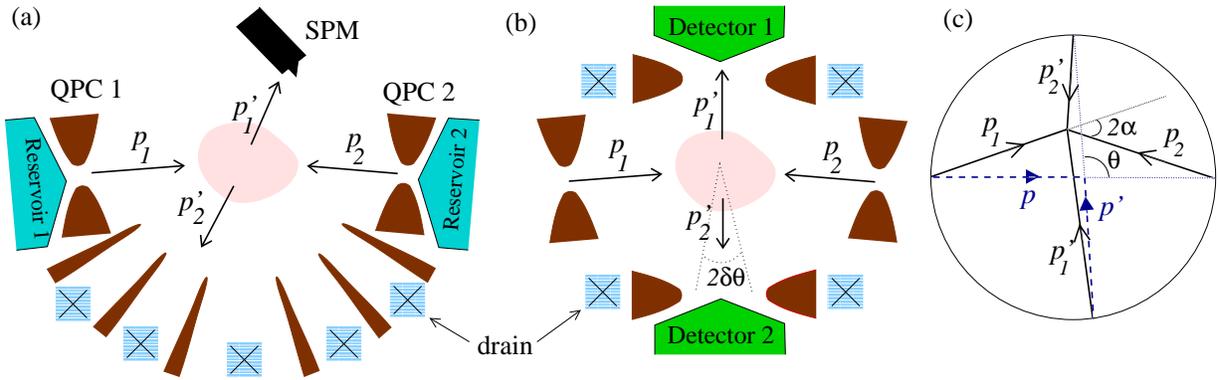}} \par}

\caption{\label{setup}{\footnotesize Setup. Two quantum point contacts (QPC)
allows the injection of electrons from two reservoirs with initial
momenta \protect\( \mathbf{p}_{1}\simeq -\mathbf{p}_{2}\protect \).
(a) Measuring the conductance as a function of the SPM tip position
gives an estimate of the electron flux} \cite{Top00} {\footnotesize and,
thereby, of the differential scattering length \protect\( \lambda (\theta )\protect \)
(top). Alternatively, one can define {}``bins'' spanning different
angles, and collect the current in the drain contacts (bottom). (b)
EPR setup. The electrons are collected in two detectors (with an aperture
angle \protect\( 2\delta \theta \protect \)) placed such that only
electrons emerging from collisions with a scattering angle around \protect\( \pi /2\protect \)
are detected. Because of antisymmetrization, the scattering amplitude
identically vanishes for the spin-triplet states, allowing only the
spin-entangled singlets (EPR pairs) to be collected. (c) Scattering
parameters. The initial} {\small (\protect\( \mathbf{p}_{1},\mathbf{p}_{2}\protect \)}{\footnotesize )
and final} {\small (\protect\( \mathbf{p}_{1}',\mathbf{p}_{2}'\protect \)}{\footnotesize )
momenta are connected by a circle of radius \protect\( p'=p\protect \)
due to energy and momentum conservation, where the relative momenta}
{\small are \protect\( \mathbf{p}=(\mathbf{p}_{1}-\mathbf{p}_{2})/2,\mathbf{p}'=(\mathbf{p}_{1}'-\mathbf{p}'_{2})/2\protect \)}
{\footnotesize and} {\small \protect\( \theta =\protect \angle (\mathbf{p},\mathbf{p}')\protect \)}
{\footnotesize is the scattering angle.} {\small The Cooper channel
is defined by \protect\( 2\alpha =\protect \angle (\mathbf{p}_{1},-\mathbf{p}_{2})\to 0\protect \).}}
\end{figure*}

We start in Section \ref{sec set} by describing  the envisioned
setup and the mechanism for the production of EPR
pairs. We write down in Sec. \ref{sec cal} the problem in a Fermi
liquid approach, solve the Bethe-Salpeter Eq. in the ladder approximation,
using RPA and considering the Cooper channel (opposite incident momenta).
The solution for the scattering amplitude (the \( t- \)matrix) is
written in Eq. (\ref{tmatrix result}) in terms of a Fourier series,
with explicit expressions for the coefficients. We study in Sec. \ref{sec res}
the scattering cross-section, and address in more details the issue
of the production and detection of the EPR pairs. 
We investigate in Sec. \ref{sec phon} the electron-electron
interaction mediated by phonons, and show that it does not have a
significant effect on the scattering.  
In Sec. \ref{kohnlut}
we show that no superconducting instabilities arises from the Kohn-Luttinger
mechanism\cite{Lut66}.
We finally consider in more detail the case of small \( r_{s} \) 
in the Appendix.
We derive in Appendix \ref{smallrsapprox}
analytical expressions for the scattering amplitude and its derivative
at \( \theta =\pi /2 \). 
In Appendix \ref{sec smallrs} we develop a
different calculation valid for very small \( r_{s} \), which is needed to
estimate the contribution of forward scattering states.

\section{Setup and production of EPR pairs\label{sec set}}

The setup for the study of Coulomb collisions in a 2DEG is described
in Fig. \ref{setup}. Two quantum point contacts (QPC) tuned in their
lowest transversal mode filter electrons escaping from two thermal
reservoirs \cite{Top00}, and allow them to collide with incident
momenta \( \mathbf{p}_{1}\simeq -\mathbf{p}_{2} \) and final momenta
\( \mathbf{p}_{1}',\mathbf{p}_{2}' \). A way to measure the scattering
length is to use an SPM tip and to record the conductance across the
sample, which provides an estimate for the local electron flux \cite{Top00};
alternatively, one can define {}``bins'' spanning different angles,
and collect the current in the drain contacts; see Fig. \ref{setup}(a).
For the production of EPR pairs, described in Fig. \ref{setup}(b),
the electrons are collected at two detectors placed so that only collisions
with a scattering angle \( \theta  \) within a small window \( \delta \theta  \)
around \( \pi /2 \) are collected: \( \theta \in [\pi /2-\delta \theta ,\pi /2+\delta \theta ] \).
We introduce the conserved total momentum \( \mathbf{P}=\mathbf{p}_{1}+\mathbf{p}_{2}=\mathbf{p}_{1}'+\mathbf{p}_{2}' \),
the relative momenta \( \mathbf{p}=\frac{1}{2}(\mathbf{p}_{1}-\mathbf{p}_{2}) \),
\( \mathbf{p}'=\frac{1}{2}(\mathbf{p}_{1}'-\mathbf{p}_{2}') \), and
the scattering angle \( \theta =\angle (\mathbf{p}',\mathbf{p}) \);
see Fig. \ref{setup}(c). The most favorable arrangement is the Cooper
channel \cite{Mah00} \begin{equation}
\label{cooperchannel}
\mathbf{p}_{2}\simeq -\mathbf{p}_{1},
\end{equation}
 as is yields conservation of the individual energies: \( p_{1}\simeq p_{2}\simeq p_{1}'\simeq p_{2}' \),
where \( p_{i}=|\mathbf{p}_{i}| \). As a consequence, the scattering
angle \( \theta \simeq \angle (\mathbf{p}_{1}',\mathbf{p}_{1}) \)
can be easily determined, while the EPR pairs have the same energy
and should therefore arrive in the detectors at the same time. We
consider incoming electrons with small excitation energies \( \xi _{i}=\hbar ^{2}p_{i}^{2}/2m-E_{F}\ll E_{F} \)
above the Fermi energy \( E_{F}=\hbar ^{2}k_{F}^{2}/2m \) of the
2DEG (\( m \) is the effective mass)\cite{Note momenta}. 

Now we describe in more details the production of spin-entangled electrons.
First, we use the fact that the two-particle interference is totally
destructive for \emph{}fermions colliding with a scattering angle
\( \theta =\pi /2 \), while, in contrast, the scattering of bosons
is enhanced compared to the classical value. This is seen in the corresponding
cross sections \( \sigma _{B/F}=|f(\theta )\pm f(\pi -\theta )|^{2} \).
Second, the fermionic character of a pair of particles also depends
on its spin-state \cite{Tay72,Bur00,Jac01}: a spin-singlet electron
pair \begin{equation}
\label{ll}
\left| S\right\rangle =(\ket {\up \down }-\ket {\down \up })/\sqrt{2}
\end{equation}
 behaves, in a spin-independent collision, like a bosonic pair because
of its symmetrical orbital wavefunction, while the triplets \begin{equation}
\label{ll}
\ket {T_{0}}=(\ket {\up \down }+\ket {\down \up })/\sqrt{2}\, \, ;\, \, \ket {T_{\pm }}=\ket {\up \up },\ket {\down \down }
\end{equation}
 behave like fermions. It is then clear that a \( \pi /2 \) scattering
experiment could distill the singlet part of uncorrelated pairs of
electrons --at least in free space.

It might seem surprising to be able to collect spin singlet states
when starting from two electrons having no spin correlations (they
come from two independent unpolarized reservoirs). However, a spin-singlet
component is always present, as seen from the change of basis for
the density matrix describing the two spin state: \begin{eqnarray}
\rho  & = & \frac{1}{4}\bbbone =\frac{1}{4}\sum _{\sigma ,\sigma '=\up ,\down }\left| \sigma \sigma '\right\rangle \left\langle \sigma \sigma '\right| \nonumber \\
 & = & \frac{1}{4}\left| S\right\rangle \left\langle S\right| +\frac{1}{4}\sum _{\mu =0,\pm }\left| T_{\mu }\right\rangle \left\langle T_{\mu }\right| ,\label{tripsingdens} 
\end{eqnarray}
 where \( \ket {\sigma \sigma '} \) corresponds to the two-electron state where
the electron injected from the first (second) reservoir has spin \( \sigma  \)
(\( \sigma ' \)).

A real detector has a small, but finite aperture angle \( 2\delta \theta  \)
around \( \theta =\pi /2 \), so that triplets will always be present.
To examine the efficiency of this collision entangler, we will define
in Sec. \ref{sec res} the ratio \( \mathcal{R} \) between the number
of scattered triplets and singlets, \( N_{T/S} \). We will find\begin{equation}
\label{ll}
\mathcal{R}=\frac{N_{T}}{N_{S}}
\simeq \delta \theta ^{2}\left| \frac{f'(\pi /2)}{f(\pi /2)}\right| ^{2},
\end{equation}
 and show that \( \left| f'/f\right| ^{2}\sim 1 \) at \( \theta =\pi /2 \).
Therefore, the number of triplets (which we want to avoid as they
can be in an unentangled product state $| T_\pm \rangle$) is negligible for small \( \delta \theta  \).

\section{Calculation of the scattering amplitude\label{sec cal}}

We study in this section the scattering between two electrons that
are both above the Fermi surface ---as opposed to  the standard
calculation of the electron lifetime due to scattering of one electron
above the surface with all the electrons present below the surface
\cite{Mah00,Ste67}. We consider a clean 2D Fermi liquid with Coulomb interaction,
neglecting impurity scattering (the mean free path can be around \( 10\, \mu \mathrm{m} \),
which is larger than the size \( L\simeq 1\, \mu \mathrm{m} \) of
the envisioned setup \cite{Top00}). The effect of phonons will be
considered in Sec. \ref{sec phon}.

\subsection{RPA and Bethe-Salpeter equation}

In 2D, the scattering amplitude \( f \) for two particles with a
relative momentum \( \mathbf{p} \) is linked to the \( t- \)matrix
via the relation \cite{Bar83} \begin{equation}
\label{scatamp}
f(\theta )=-\frac{m}{\hbar ^{2}\sqrt{2\pi p}}\, \, t(\theta ),
\end{equation}
 with the corresponding scattering cross-section ({}``length'' in
2D) \( \lambda (\theta )=|f(\theta )|^{2} \). The 2D Coulomb interaction
in vacuum is \cite{3dlines,And82} \begin{equation}
\label{vbaree}
V_{C}(\mathbf{q})=\int d\mathbf{r}e^{-i\mathbf{q}\cdot \mathbf{r}}V_{C}(r)=\frac{2\pi e_{0}^{2}}{q},
\end{equation}
 where \( e_{0}^{2}=e^{2}/4\pi \epsilon _{0}\epsilon _{r} \). We
included the dielectric constant \( \epsilon _{r} \) of GaAs and
set \( p=k_{F} \) for future comparisons with the scattering in GaAs.
The corresponding exact scattering \emph{t-}matrix is given by \cite{Bar83}
\begin{equation}
\label{tbare}
t_{\mathrm{C}}(\theta )=\frac{\varsigma }{\sin |\theta /2|}\frac{\hbar ^{2}\sqrt{\pi }}{m}\frac{\Gamma \left( \frac{1}{2}+i\varsigma \right) }{\Gamma \left( 1-i\varsigma \right) }e^{i\pi /4-2i\varsigma \ln |\sin \theta /2|},
\end{equation}
with \( \Gamma (x) \) the Gamma function and \( \varsigma =me_{0}^{2}/k_{F}\hbar ^{2} \).
This yields\begin{equation}
\label{tbarenophase}
|t_{\mathrm{C}}(\theta )|=\frac{\hbar ^{2}\sqrt{\pi \varsigma \tanh (\pi \varsigma )}}{m\sin |\theta /2|}.
\end{equation}
Note that $|t_C|\neq V_C$, contrary to the situation in $3D$.

\begin{figure}[t]
{\centering \resizebox*{0.9\columnwidth}{!}{\includegraphics{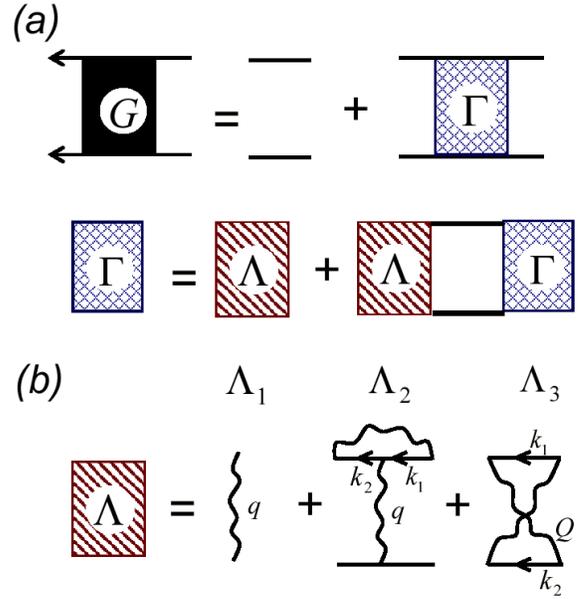}} \par}

\caption{\label{diagram}{\small (a) The two-particle Green function and the
Bethe-Salpeter equation for the vertex \protect\( \Gamma \protect \).
We only show the direct diagrams (i.e., without the exchange ones).
(b) Lowest order diagrams (\protect\( \Lambda _{1,2,3}\protect \))
contributing to the irreducible vertex \protect\( \Lambda \protect \).
The wavy lines denote the screened Coulomb interaction \protect\( V\protect \),
given in RPA by resuming the bubble diagrams.}}
\end{figure}

Next we include the effect of the many-particle background, and calculate
the \emph{t-}matrix in the presence of the Fermi sea. The \emph{t-}matrix
is given by the vertex function \( t=\Gamma (\omega _{1,2}\to \xi _{1,2}) \)
appearing in the two-particle Green function \cite{FW71}; see Fig.
\ref{diagram}(a). We note that the arrangement \( \mathbf{p}_{2}=-\mathbf{p}_{1} \)
corresponds to the well-known Cooper channel, discussed for instance
in the context of Cooper instability \cite{Mah00}. Consequently,
we adopt the approach used by Kohn and Luttinger in their work on
intrinsic superconductivity in a 3D Fermi liquid \cite{Lut66}. However,
our calculation differs in two ways. First, we consider a 2D system
where the screened potential is non-analytic (because of the modulus
\( q=|\mathbf{q}| \) instead of \( q^{2} \) in 3D). Secondly, we
are interested in the scattering amplitude, while Kohn and Luttinger
focused on the instability in the vertex arising from spherical harmonics
of the crossed ('exchange') diagram (see \( \Lambda _{3} \) below). 

The two-particle Green function in real space is \begin{equation}
\label{i}
G_{\sigma _{1}'\sigma _{2}'\sigma _{1}\sigma _{2}}(1',2';1,2)=(-i)^{2}\left\langle \mathcal{T}c_{1'\sigma _{1}'}c_{2'\sigma _{2}'}c^{\dagger }_{1\sigma _{1}}c_{2\sigma _{2}}^{\dagger }\right\rangle ,
\end{equation}
with the notation \( i=(\mathbf{x}_{i},t_{i}) \), \( \sigma _{i}=\, \up ,\, \down  \)
are the spin indices, and \( \mathcal{T} \) the time-ordering operator.
For a spin-independent Hamiltonian we can write \cite{FW71,Lut66}
\begin{eqnarray}
G_{\sigma _{1}'\sigma _{2}'\sigma _{1}\sigma _{2}}(1',2';1,2) & = & G(1)G(2)(2\pi )^{3}\nonumber \label{i} \\
 &  & \hspace {-30mm}\times \left[ \delta (1'-1)\delta _{\sigma _{1}'\sigma _{1}}\delta _{\sigma _{2}'\sigma _{2}}-\delta (1'-2)\delta _{\sigma _{1}'\sigma _{2}}\delta _{\sigma _{2}'\sigma _{1}}\right] \nonumber \\
 &  & \hspace {-30mm}+(i/\hbar )G(1')G(2')\Gamma _{-}(1',2';1,2)G(1)G(2),\label{gf2part} 
\end{eqnarray}
with the single-particle Green function \( G(i) \) and the (anti-)symmetrized
vertex \( \Gamma _{\pm }(1',2';1,2)=\Gamma (1',2';1,2)\pm \Gamma (1',2';2,1) \)
expressed in terms of the unsymmetrized \( \Gamma  \). We consider
the singlet/triplet basis \( \{\ket {S},\ket {T_{0}},\ket {T_{\pm }}\} \)
and introduce the corresponding creation operators\begin{eqnarray}
a^{\dagger }_{S/T_{0}}(1,2) & = & 1/\sqrt{2}\left( c_{1\up }^{\dagger }c_{2\down }^{\dagger }\mp c_{1\down }^{\dagger }c_{2\up }^{\dagger }\right) ,\\
a^{\dagger }_{T_{\pm }}(1,2) & = & c_{1\up }^{\dagger }c_{2\up }^{\dagger }\, \, ,\, \, c_{1\down }^{\dagger }c_{2\down }^{\dagger },\label{i} 
\end{eqnarray}
 which we use to define the related two-particle Green functions\begin{eqnarray}
\hspace {-5mm}G_{S/T_{0}}(1',2';1,2) & = & -\left\langle \mathcal{T}a_{S/T_{0}}(1',2')a_{S/T_{0}}^{\dagger }(1,2)\right\rangle \nonumber \label{1} \\
 & \hspace {-15mm}= & \hspace {-8mm}\frac{1}{2}\left\{ G_{\up \down \up \down }+G_{\down \up \down \up }\mp G_{\up \down \down \up }\mp G_{\down \up \up \down }\right\} ;\\
G_{T_{\pm }}(1',2';1,2) & = & -\left\langle \mathcal{T}a_{T_{\pm }}(1',2')a_{T_{\pm }}^{\dagger }(1,2)\right\rangle \nonumber \\
 & = & G_{\up \up \up \up ,\down \down \down \down }\, \, \, \, ,
\end{eqnarray}
where we have dropped the argument \( (1',2';1,2) \) for ease of
notation. Using (\ref{gf2part}), we find that the spin and orbital
symmetry of the pair of particles is directly reflected in the vertex
\( \Gamma  \):\begin{eqnarray}
G_{S/T}(1',2';1,2) & = & -G(1)G(2)\left[ \delta (1'-1)\pm \delta (1'-2)\right] (2\pi )^{3}\nonumber \label{i} \\
 &  & \hspace {-15mm}+(i/\hbar )G(1')G(2')\Gamma _{\pm }(1',2';1,2)G(1)G(2),\label{Gst} 
\end{eqnarray}
 where \( T \) denotes one of the triplet states \( T_{0,\pm } \).
From Eq. (\ref{Gst}) we see that the vertex (and therefore the scattering
amplitude) has either a bosonic (symmetric) or fermionic (antisymmetric)
behavior depending on the spin state. Therefore, we calculate the
\emph{unsymmetrized} vertex \( \Gamma  \) \emph{}giving the \emph{t-}matrix
\( t \) and the scattering amplitude \( f \), and (anti)symmetrize
the latter according to the spin state.

From now on we consider the Green function in momentum and frequency
space. Taking into account conservation of momentum and frequency,
the vertex satisfies the Bethe-Salpeter equation \cite{FW71,Lut66}
\begin{eqnarray}
\Gamma (\tilde{p}',\tilde{p};\tilde{P}) & = & \Lambda (\tilde{p}',\tilde{p};\tilde{P})\nonumber \\
 &  & \hspace {-20mm}+\frac{i}{\hbar (2\pi )^{3}}\int d\tilde{k}\Lambda (\tilde{k},\tilde{p};\tilde{P})G(\tilde{k}_{1})G(\tilde{k}_{2})\Gamma (\tilde{p}',\tilde{k};\tilde{P}),\label{BSstart} 
\end{eqnarray}
illustrated in Fig. \ref{diagram}(a). We have introduced the irreducible
vertex \( \Lambda  \), the intermediate momenta \( \tilde{k}_{1,2}=\tilde{P}/2\pm \tilde{k} \)
given in terms of the relative momentum \( \tilde{k} \), the frequency
\( \omega  \) and the notation \( \tilde{p}=(\mathbf{p},\omega ) \).
All the possible exchange diagrams should be included in \( \Lambda  \)
while avoiding double-counting physically equivalent diagrams. We
first consider zero temperature \( k_{B}T=0 \), and discuss finite
\( T \) effects later. 

In a first stage, we use the random phase approximation (RPA) for
the many-electron background \cite{FW71}; this yields the screened
interaction \begin{equation}
\label{v2rpa}
V(\tilde{q})=\frac{V_{C}(q)}{1-V_{C}(q)\chi ^{0}(\tilde{q})},
\end{equation}
 given in terms of the bubble susceptibility \( \chi ^{0} \), with
the momentum transfer \( \tilde{q}=(\mathbf{q},\omega _{q})=\tilde{p}'-\tilde{p} \)
. The RPA requires a high density, which is controlled in 2D via the
parameter \( r_{s}=me_{0}^{2}/(\hbar ^{2}\sqrt{\pi n})\ll 1 \), where
\( n \) is the electronic sheet density. First, one can consider
the static limit \cite{Ste67} \begin{equation}
\label{pol2d}
\chi ^{0}(q,\omega _{q}=0)=-\frac{m}{\pi \hbar ^{2}}\left[ 1-\Theta (q>2k_{F})\sqrt{1-\left( \frac{2k_{F}}{q}\right) ^{2}}\right] 
\end{equation}
 because the dependence of \( \chi ^{0} \) with \( \omega _{q} \)
is smooth, and can be therefore neglected in the integration of intermediate
fermionic lines ---which, as we shall see below, selects only intermediate
states \( \mathbf{k} \) at the Fermi surface: \( \xi _{k}=0\then \hbar \omega _{q}=\xi _{k}-\xi _{1}=0 \)
for \( \xi _{1}=0 \). Note that the divergence in the \( \omega _{q} \)---integration
of \( V(\tilde{q})\sim 1/(\omega _{q}-\omega _{p}) \) near the plasmon
frequency \( \omega _{p} \) disappears because it is an odd function
of \( \omega  \). Next, the \( q \)--dependent part of \( \chi ^{0}(q,\omega _{q}=0) \)
in 2D vanishes at the Fermi surface, as \( q\simeq 2k_{F}\sin |\theta /2|<2k_{F} \). This
justifies the standard Thomas-Fermi screening\begin{equation}
\label{v2d}
V(q)=\frac{2\pi e_{0}^{2}}{q+k_{s}},
\end{equation}
 with the screening momentum \( k_{s}=2me_{0}^{2}/\hbar ^{2}=k_{F}r_{s}\sqrt{2} \). 

Within RPA, the renormalized one-particle Green function is given
by\cite{Mah00}: \begin{equation}
\label{ll}
G(\tilde{k})\simeq \frac{Z}{\omega _{k}-\xi _{k}^{*}-i\, \imag \Sigma (\tilde{k})}
\end{equation}
for small \( \omega _{k} \) and \( \xi _{k} \). In 2D, for GaAs
and \( k_{B}T,\xi _{k}\to 0 \), one has the renormalization factor\cite{Bur00,Gal04}
\( Z=1-r_{s}(1/2-1/\pi )/\sqrt{2}\simeq 0.62 \), the renormalized
mass \cite{Gal04} \( m^{*}=m[1-\ln (1/r_{s})r_{s}/\pi ]\simeq 0.96\, m \)
entering in \( \xi _{k}^{*} \), and the broadening (inverse lifetime)
\cite{Giu82} \( \imag \Sigma \sim \xi ^{2}\ln \xi  \). The latter
vanishes for particles near the Fermi surface (\( \xi _{k}\to 0 \)),
which corresponds to well-defined quasi-particle states. For simplicity\cite{renorma},
we set \( Z=1 \), \( m^{*}=m \), and therefore approximate the renormalized
Green function by the free propagator\begin{equation}
\label{a}
G(\tilde{k})\simeq G_{0}(\tilde{k})=\frac{1}{\omega _{k}-\xi _{k}}.
\end{equation}

We now consider the irreducible vertex \( \Lambda  \) \emph{}in lowest
orders in \( V \),as 
\( m V/\pi \hbar^2\sim me^{2}_{0}/k_{F}\sim r_{s}\ll 1 \).
The lowest-order diagrams are shown in Fig. \ref{diagram}(b); they
are the single interaction line \( \Lambda _{1} \), the vertex renormalization
\( \Lambda _{2} \), and the crossed diagram \( \Lambda _{3} \):\begin{equation}
\label{lam1}
\Lambda _{1}=V(q),
\end{equation}
\begin{equation}
\label{lam2}
\Lambda _{2}=V(q)\frac{1}{(2\pi )^{2}}\int d\mathbf{k}_{1}B(\mathbf{k}_{1},\tilde{q})\left[ V(\mathbf{k}_{1}-\mathbf{p}_{1})+V(\mathbf{p}_{2}'-\mathbf{k}_{1})\right] ,
\end{equation}
\begin{equation}
\label{lam3}
\Lambda _{3}=\frac{1}{(2\pi )^{2}}\int d\mathbf{k}_{1}B(\mathbf{k}_{1},\tilde{Q})V(\mathbf{k}_{1}+\mathbf{p}_{2}')V(\mathbf{k}_{1}+\mathbf{p}_{2}),
\end{equation}
with \( \tilde{q}=\tilde{p}_{1}'-\tilde{p}_{1} \) and \( \tilde{Q}=\tilde{p}_{2}'-\tilde{p}_{1} \).
The function \begin{equation}
\label{bfun}
B(\mathbf{k},\tilde{q})=\frac{n(\mathbf{k}+\mathbf{q})-n(\mathbf{k})}{\xi _{\mathbf{k}+\mathbf{q}}-\xi _{\mathbf{k}}-\hbar \omega \pm i\eta }
\end{equation}
 arises from the frequency integration of the bubble diagram, and
involves the Fermi occupations factors \( n(\mathbf{k})=\Theta (-\xi _{k}) \)
 at $k_B T=0$.
We can estimate \( \Lambda _{2}\simeq V(q)V(k_{F})\int d\mathbf{k}_{1}B(\mathbf{k}_{1},\tilde{q})/(2\pi )^{2}=V(q)V(k_{F})\chi ^{0}(\tilde{q})\simeq -V(q)k_{s}/k_{F} \)
(apart from negligibly small integration regions), and similarly
for \( \Lambda _{3} \). 
Hence \cite{Chu89}, \( \Lambda _{2,3}/\Lambda _{1}\sim r_{s}\ll 1 \)
and we can keep only the direct interaction line \( \Lambda \simeq \Lambda _{1}=V \),
which corresponds to the \emph{ladder approximation}. The criterion
for the validity of the ladder approximation is usually expressed \cite{FW71}
by \( \lambda k_{F}\ll 1 \). This low density regime is nevertheless
consistent with RPA, as can be seen, e.g., from the Born approximation
\begin{equation}
\label{rpaladdcheck}
t\simeq V_{C}\then \lambda k_{F}\sim \frac{1}{2\pi }\left[ \frac{mV_{C}(k_{F})}{\hbar ^{2}}\right] ^{2}\sim \left( \frac{k_{s}}{k_{F}}\right) ^{2}\ll 1.
\end{equation}
In summary, we need to solve the following Bethe-Salpeter equation:\begin{eqnarray}
\hspace {-5mm}\Gamma (\tilde{p}'-\tilde{p};\tilde{P}) & = & V(p'-p)\nonumber \\
 &  & \hspace {-25mm}+\frac{i}{\hbar (2\pi )^{3}}\int d\tilde{k}V(\tilde{k}-\tilde{p})G_{0}(\tilde{k}_{1})G_{0}(\tilde{k}_{2})\Gamma (\tilde{p}'-\tilde{k};\tilde{P}),\label{BS} 
\end{eqnarray}
where \( \tilde{k}_{1,2}=\frac{1}{2}\tilde{P}\pm \tilde{k}. \)

\subsection{Energy integration and logarithmic factor}

We now follow the derivation of the Cooper instability \cite{Mah00},
including a discussion of the less standard case of particles that
are not in the Cooper channel, i.e. with \( \mathbf{p}_{1}\neq -\mathbf{p}_{2} \).
Before solving Eq. (\ref{BS}) to all orders, we first consider its
second-order iteration, \( \Gamma ^{(2)}(\mathbf{p}'-\mathbf{p};\tilde{P})=V(\mathbf{p}'-\mathbf{p})+i/\hbar (2\pi )^{3}\, \int d\tilde{k}V(\mathbf{k}-\mathbf{p})G(\tilde{k}_{1})G(\tilde{k}_{2})V(\mathbf{p}'-\mathbf{k}) \).
The \( \omega _{k} \)-integration of the Green functions yields \cite{FW71}\begin{eqnarray}
D(\mathbf{k}_{1},\mathbf{k}_{2}) & := & \frac{i}{2\pi \hbar }\int d\omega _{k}G_{0}(\mathbf{k}_{1},\frac{\Omega }{2}+\omega _{k})G_{0}(\mathbf{k}_{2},\frac{\Omega }{2}-\omega _{k})\nonumber \\
 &  & \hspace {-5mm}=\frac{N(k_{1},k_{2})}{\xi _{1}+\xi _{2}-\xi _{k_{1}}-\xi _{k_{2}}+2i\eta N(k_{1},k_{2})},\label{eq2} 
\end{eqnarray}
 with the function \( N(k_{1},k_{2}):=1-n(k_{1})-n(k_{2}) \). The
\( \tilde{P} \)-frequency \( \hbar \Omega  \) has been set to \( \xi _{1}+\xi _{2} \)
to retrieve the \emph{t-}matrix, and \( \eta  \) can be set to \( 0 \)
in the denominator. We now consider the Cooper channel \( \mathbf{p}=\mathbf{p}_{1}=-\mathbf{p}_{2},\mathbf{k}=\mathbf{k}_{1}=-\mathbf{k}_{2} \),
which gives \( q=|\mathbf{p}'-\mathbf{p}|=2p\sin |\theta /2| \),
and \( n(k_{1})=n(k_{2})=\Theta (-\xi _{k})\then N(k)=\mathrm{sgn}(\xi _{k}) \).
This yields a single discontinuity in the numerator when \( \xi _{k}=0 \),
which coincides with the zero of the denominator (\( \xi _{k}=\xi  \))
when considering a vanishing excitation energy for the incident particles:
\( \xi =\xi _{1,2}\to 0 \).

As a consequence, the main contribution to the energy integration
comes from virtual states at the Fermi surface, i.e. \( \xi _{k}\simeq 0 \).
We set \( k=k_{F} \) in \( V \), and integrate only on \( D(k)=D(\mathbf{k},\mathbf{k}) \).
The dominant term comes from both sides around \( k=k_{F} \) and
yields the factor \cite{linear}\begin{equation}
\label{log-div}
\nu :=\frac{1}{2\pi }\int ^{\infty }_{0}dk\, k\, D(k)\simeq \frac{m}{2\pi \hbar ^{2}}\log \frac{\xi }{E_{F}}.
\end{equation}
This logarithmic divergence (as \( \xi \to 0 \)) plays a crucial
role here. It allows us to neglect the \( k- \)dependence of \( V \),
as it selects only intermediate states at the Fermi surface (their
contribution is logarithmically dominant as \( \xi \to 0 \)). This
is the main effect of the fermionic background on the scattering,
apart from the screening.

We now consider finite temperatures with \( n(k)=(1+e^{\xi _{k}/k_{B}T})^{-1} \).
For \( \xi \sim k_{B}T\ll E_{F} \), we find
\begin{eqnarray}
\nu (T)\simeq \frac{m}{4\pi \hbar ^{2}}\left\{ \log \frac{\xi ^{2}-(k_{B}T)^{2}}{E^{2}_{F}}\right.  &  & \nonumber \\
 &  & \hspace {-35mm}\left. +\frac{\xi }{k_{B}T}\left[ \log \frac{k_{B}T-\xi }{k_{B}T+\xi }-\pi i\right] +2\right\} .\label{1} 
\end{eqnarray}
For \( \xi \gg k_{B}T \) we recover (\ref{log-div}), while for \( \xi \ll k_{B}T \)
we find \begin{equation}
\label{nut}
\nu (T)\simeq \frac{m}{2\pi \hbar ^{2}}\log \frac{k_{B}T}{E_{F}}.
\end{equation}
This logarithmic factor (\ref{nut}) is the 2D equivalent of the one
found in the discussion of the Cooper instability for both phonon-mediated
\cite{FW71} or Kohn-Luttinger \cite{Lut66} superconductivity ---with
the 2D density of state \( \nu _{2D}=m/(2\pi \hbar ^{2}) \) instead
of the 3D one. 

Different cut-offs arise if the Cooper channel condition (\( P=0 \))
is not strictly respected. For instance, experiments might require
a small but finite angle \( 2\alpha =\angle (\mathbf{p}_{1},-\mathbf{p}_{2})\ll 1 \)
between the incident particles to prevent misalignment, in which case
we have a total momentum \( P=2p_{1}\sin \alpha \simeq 2k_{F}\alpha  \)
if \( p_{1}=p_{2} \) and \( |p_{1}-k_{F}|,P\ll k_{F} \). Alternatively,
particles might be injected in perfect opposite direction (\( \alpha =0 \)),
but with a different energy (this can arise e.g. in case of hot electrons,
see Sec.\ref{hot}), i.e. \( p_{1}\neq p_{2} \), leading to \( P=|p_{1}-p_{2}| \).
In these cases, we find at \( T=0 \) \begin{equation}
\label{nnotcooper}
N(k,\phi )=\protect {\Theta (k-k_{+})}-\Theta (k_{-}-k),
\end{equation}
 where \( k_{\pm }=\sqrt{k_{F}^{2}-\left( P\sin \phi \right) ^{2}}\pm P\cos \phi  \),
and \( \phi  \) is the integration angle for \( \mathbf{k} \) {[}\( \phi =\angle (\mathbf{k},\mathbf{p})+\pi /2 \)
if \( \alpha \neq 0 \); \( \phi =\angle (\mathbf{k},\mathbf{p}) \)
if \( p_{1}\neq p_{2} \){]}. In the limit \( |p_{1,2}-k_{F}|\ll P \)
we find\begin{equation}
\label{nua}
\nu (\phi )\simeq \frac{m}{2\pi \hbar ^{2}}\log \left( \frac{P}{k_{F}}\sin |\phi |\right) ,
\end{equation}
where we recall that \( P=2k_{F}\alpha  \) or \( |p_{1}-p_{2}| \). 

Therefore, the divergence (due to the single discontinuity of the
two Fermi surfaces for particles in the Cooper channel) is cut-off
by \( \max \{\xi /E_{F},k_{B}T/E_{F},P/k_{F}\} \). Away from the
Cooper channel (\( P\simeq k_{F} \)), or for large temperatures (\( k_{B}T,\xi \simeq E_{F} \))
the logarithmic factor disappears; in that case the fact that only
virtual states having the Fermi energy (\( k\simeq k_{F} \)) contribute
to higher orders does not apply. When \( p_{1,2}\simeq k_{F} \),
we must make sure that the ingoing and outgoing states are available,
i.e. \( p_{1,2},p'_{1,2}>k_{F} \). For \( \pi /2 \)-scattering and
\( \alpha \neq 0 \), we have \( p'_{1,2}=p_{1}(\cos \alpha \pm \sin \alpha )\then  \)\( p_{1}\alt k_{F}(1+\alpha ) \)
{[}see Fig. \ref{setup}(c){]}; then \( \alpha =P/2k_{F}<\xi /E_{F} \).
Similarly for \( p_{1}\neq p_{2} \), we need \( p_{2}=p_{1}-P>k_{F}\then P/2k_{F}<\xi _{1}/E_{F} \).
As a consequence, the cut-off is determined by either
\( \xi  \) or \( k_{B}T \).

\begin{figure}[t]
{\centering \resizebox*{0.95\columnwidth}{!}{\includegraphics{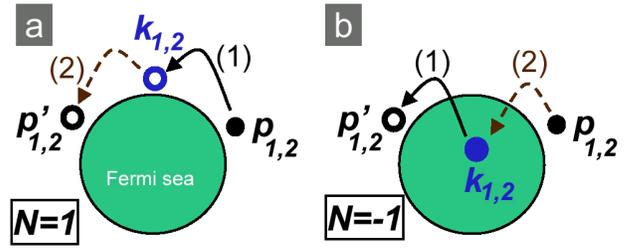}} \par}

\caption{\label{manypart}Scattering with a fermionic background. (a) Direct
virtual transition: the initial states \protect\( p_{1,2}\protect \)
first go to the available intermediate state \protect\( k_{1,2}>k_{F}\protect \)
(1), and then go to the final states \protect\( p_{1,2}'\protect \)
(2). This process is represented by a term \protect\( c^{\dagger }_{p'}c_{k}c^{\dagger }_{k}c_{p}\protect \).
(b) Exchange process corresponding to \protect\( c^{\dagger }_{p'}c^{\dagger }_{k}c_{k}c_{p}\protect \):
intermediate states with \protect\( k_{1,2}<k_{F}\protect \) first
fill the final states \protect\( p'_{1,2}\protect \), creating a
particle-hole excitation (1); the holes are subsequently filled by
the initial states \protect\( p_{1,2}\protect \) (2). The effect
of the many-particle fermionic background manifests itself in the
function \protect\( N(k_{1},k_{2})\protect \) in Eq. (\ref{eq2}),
which adds a negative sign for the exchange process, and is therefore
responsible, in the Cooper channel, for the logarithmic term (\ref{log-div})
that selects states at the Fermi surface, \protect\( k_{1,2}\simeq k_{F}\protect \).
In vacuum, we have only case (a), so 
\protect\( N(k_{1},k_{2})= 1\protect \) 
and all states contribute with the same sign; 
this yields no divergence, and therefore no selection.}
\end{figure}

Before proceeding with the solution of the Bethe-Salpeter Eq., we
comment on the difference with the standard scattering theory in vacuum,
where the scattering \( t- \)matrix is given by the Lippmann-Schwinger
equation\cite{Tay72} \( t^{\mathrm{vac}}_{E}=V+VG^{\mathrm{vac}}_{E}t_{E}^{\mathrm{vac}} \)
for an incoming energy \( E \). As the single-particle Green function
\( G^{\mathrm{vac}}_{E}(\mathbf{k})=1/(E-E_{k}) \) is an odd-function
around the divergence at \( E_{k}=E \), no divergence develops, and
no selection of intermediate states \( \mathbf{k} \) occurs.

In the case of a Fermi sea, the Lippmann-Schwinger Eq. is replaced
by the Bethe-Salpeter Eq. (which can be written symbolically \( \Gamma =V+VD\Gamma  \)),
and \( G^{\mathrm{vac}}_{E} \) corresponds to the factor \( D \),
Eq. (\ref{eq2}). This is seen by setting \( E_{F}=0 \), giving \( N(k_{1},k_{2})\to 1 \)
and thus \( D\to G_{E}^{\mathrm{vac}}(k) \) for the relative momentum
\textbf{\( \mathbf{k}=(\mathbf{k}_{1}-\mathbf{k}_{2})/2 \)}. The
difference between the Fermi sea and the vacuum cases lies in the
numerator \( N(k_{1},k_{2}) \), which is a direct effect of Fermi
statistics and assigns a negative sign to the exchange processes where
the transitions occur via two states \( k_{1},k_{2}<k_{F} \) below
the Fermi surface, as compared to the direct processes via intermediate
states above the Fermi surface, \( k_{1,2}>k_{F} \) (see Fig. \ref{manypart}).
In the Cooper channel, this factor is responsible for the selection
of intermediate states at the Fermi energy (\( k_{1,2}\simeq k_{F} \)),
via the log-divergence (\ref{log-div}); the latter arises because
\( N(k)=\mathrm{sgn}(\xi _{k}) \) implies that \( D(\xi _{k})\sim 1/|\xi _{k}| \)
is an even function (after setting \( \xi =0 \)). We emphasize that
the selection of virtual states at the Fermi energy (\( k\simeq k_{F} \))
disappears away from the Cooper channel; see Eq. (\ref{nua}).

\subsection{Solution as a Fourier series}

We can repeat the integration over the frequency and energy described
above at every order. This results in a new Bethe-Salpeter equation\begin{equation}
\label{newbs}
t(\theta )=V(\theta )+\nu \frac{1}{2\pi }\int d\phi V(\phi )t(\theta -\phi ),
\end{equation}
with the screened 2D Coulomb potential at the Fermi surface\begin{equation}
\label{coulombfs}
v(\phi )=\frac{2\pi e^{2}}{2k_{F}\sin |\phi /2|+k_{s}}.
\end{equation}
To solve this integral equation, we expand \( v \) in a Fourier series:\begin{equation}
\label{fourdef}
v(\phi )=\sum ^{\infty }_{n=-\infty }v_{n}e^{in\phi }\, ,\, v_{n}=\frac{1}{2\pi }\int ^{\pi }_{-\pi }d\phi V(\phi )e^{-in\phi }
\end{equation}
as well as \( t(\phi ) \). The solution of the Bethe-Salpeter equation
is then simply given by\begin{equation}
\label{tmatrix result}
t(\theta )=\sum _{n}\frac{v_{n}}{1-\nu v_{n}}e^{in\theta }.
\end{equation}
This expression for the Coulomb scattering \emph{t-}matrix of two
electrons in the Cooper channel (\textbf{\( \mathbf{p}_{2}=-\mathbf{p}_{1} \)}),
in the presence of a Fermi sea, is the main result of the paper. We
note that the procedure followed here is not valid for very small
\( r_{s}\ll \xi /E_{F},k_{B}T/E_{F} \), which will be addressed later
in App.\ref{sec smallrs}.

\subsection{Fourier coefficients of \protect\( v(\phi )\protect \)}

For the Fourier coefficients of \( v(\phi ) \), we integrate (\ref{fourdef})
in the complex plane with \( z=e^{i\phi /2} \). We find\begin{equation}
\label{vnsum}
v_{n}=\frac{4e_{0}^{2}}{k_{F}\cos \gamma }\sum ^{\infty }_{\mathrm{odd}\, m\geq 1}\frac{\cos (m\gamma )}{2n+m}.
\end{equation}
with\begin{equation}
\label{gammars}
\sin \gamma =\frac{k_{s}}{2k_{F}}=\frac{r_{s}}{\sqrt{2}}.
\end{equation}
For numerical estimates, it is more convenient to write the result as
\begin{equation}
\label{2}
v_{n}=-\frac{2e_{0}^{2}}{k_{F}\cos \gamma }\left\{ \mathcal{L}_{n}+\mathcal{A}_{n}\right\} ,
\end{equation}
\begin{equation}
\label{2}
\mathcal{L}_{n}=\ln \left[ \tan \left( \frac{\gamma }{2}\right) \right] \cos (2n\gamma )-\frac{\pi }{2}\sin (2n\gamma ),
\end{equation}
\begin{equation}
\label{2}
\mathcal{A}_{n}=2\sum ^{2n-1}_{\mathrm{odd}\, m\geq 1}\frac{\cos \left[ \gamma (2n-m)\right] }{m}.
\end{equation}
For instance, for \( n=0 \) we have\begin{equation}
\label{v0}
v_{0}=-\frac{2e_{0}^{2}}{k_{F}\cos \gamma }\ln \left[ \tan \left( \frac{\gamma }{2}\right) \right] .
\end{equation}

\subsubsection{Integral approximation.}

For small \( r_{s} \) the sum (\ref{vnsum}) is smooth and can be
approximated by an integral, giving
\begin{eqnarray}
v_{n}\simeq \frac{2e_{0}^{2}}{k_{F}\cos \gamma }\left\{ \sin (2n\gamma )\left[ \frac{\pi }{2}-\mathrm{Si}(2n\gamma +\gamma )\right] \right.  &  & \nonumber \\
-\cos (2n\gamma )\mathrm{Ci}(2n\gamma +\gamma )\Big \}, &  & \label{sinint} 
\end{eqnarray}
where \( \mathrm{Si} \) and \( \mathrm{Ci} \) are the sine and cosine
integrals. This expression can also be obtained (for large \( n \))
with a linear expansion of the sine in \( v(\phi )\simeq 2\pi e_{0}^{2}/(k_{s}+k_{F}|\phi |) \)
before calculating the Fourier coefficients. For realistic parameters
it is very \emph{}accurate already for \( n\geq 1 \). It yields the
asymptotics for \( n\gg 1 \) 
\begin{equation}
\label{vnasym}
v_{n}\simeq \frac{e_{0}^{2}}{2k_{F}\gamma ^{2}\cos \gamma }\frac{1}{n^{2}}.
\end{equation}
 Hence, the large \( n \) dependence is polynomial, \( v_{n}\sim n^{-2} \),
which reflects the fact that the potential \( v(\phi ) \) is non-analytic.
This is in contrast with the 3D case, where the coefficients of the
spherical harmonics decomposition are \cite{Lut66} \( (2\pi e_{0}^{2}/k_{F}^{2})Q_{l}(1+r_{s}^{1/2}2^{-1/4}), \)
and their decay is exponential in \( l \) (\( Q_{l} \) is the Legendre
function of the second kind).

\subsubsection{Small \protect\( r_{s}\protect \) approximation.}

We now expand (\ref{sinint}) in small \( r_{s} \), and find \begin{equation}
\label{vnap1}
v_{n}\simeq -\frac{2e_{0}^{2}}{k_{F}}\left\{ \ln (2n\gamma )+\frac{1}{2}-\pi n\gamma \right\} \to -\frac{2e_{0}^{2}}{k_{F}}\ln (2n\gamma ).
\end{equation}
This expression is not valid for very large \( n \), as we expanded
to lowest order in \( n\gamma  \). It remains finite in the limit
\( r_{s}\to 0 \), because \( e_{0}^{2}\sim r_{s} \).

\section{Scattering length and EPR pairs\label{sec res}}

We now apply our result to a realistic GaAs 2DEG, and study the dependence
of the scattering amplitude on the scattering angle, \( r_{s} \)
and temperature, before discussing the production and detection of
spin-entangled electron pairs. 

\setcounter{subsubsection}{0}

\subsection{The different scattering lengths}

We define the scattering length for singlets and triplets\begin{equation}
\label{t-singtrip}
\lambda _{S/T}(\theta )=|f(\theta )\pm f(\theta -\pi )|^{2},
\end{equation}
 following (\ref{Gst}). We recall the scattering amplitude \( f(\theta ) \)
defined in (\ref{scatamp}):\begin{equation}
\label{t-scat}
f(\theta )=\frac{m}{\hbar ^{2}\sqrt{2\pi k_{F}}}t(\theta ),
\end{equation}
with the \( t- \)matrix given by (\ref{tmatrix result}). Unpolarized
sources contain \( 1/4 \) of singlets and \( 3/4 \) of triplets
{[}see Eq. (\ref{tripsingdens}){]}, which yields the scattering length
\begin{equation}
\label{t-unpol}
\lambda (\theta )=\frac{1}{4}\lambda _{S}(\theta )+\frac{3}{4}\lambda _{T}(\theta ).
\end{equation}
We also define the scattering length \( \lambda ^{(1)} \) obtained
from the Born approximation with the amplitude\begin{equation}
\label{t1-scat}
f^{(1)}(\theta )=\frac{m}{\hbar ^{2}\sqrt{2\pi k_{F}}}v(\theta ),
\end{equation}
as well as the corresponding bare scattering lengths \( \lambda _{C} \)
and \( \lambda _{C}^{(1)} \), obtained by replacing \( t(\theta ) \)
with \( t_{C}(\theta ) \) and \( v_{C}(\theta ) \), given by Eq.
(\ref{tbare}) and (\ref{vbaree}). We point out that \( \varsigma =r_{s}/\sqrt{2}. \)

\subsection{Total scattering length}

We now take typical parameters for a 2D GaAs electron gas \cite{And82},
\( \epsilon _{r}=13.1 \), \( r_{s}=0.86 \), and a sheet density
\( n=10^{15}\, \mathrm{m}^{-2} \), and assume \( \xi <k_{B}T=10^{-2}E_{F} \)
(\( T=20 \) mK). First, we estimate the magnitude of the scattering
and calculate the total scattering length integrated over \( \pi  \):\emph{\begin{equation}
\label{scat tot ex}
\lambda _{\mathrm{tot}}=\int _{0}^{\pi }d\theta \lambda (\theta )=\frac{1}{4}\lambda ^{\mathrm{tot}}_{S}+\frac{3}{4}\lambda ^{\mathrm{tot}}_{T}=3.39\, \, \mathrm{nm}.
\end{equation}
} with\begin{equation}
\label{1}
\lambda ^{\mathrm{tot}}_{S}=7.92\, \, \mathrm{nm},\, \, \lambda ^{\mathrm{tot}}_{T}=1.88\, \, \mathrm{nm}.
\end{equation}
This is consistent with the ladder approximation, which requires \( \lambda _{\mathrm{tot}}k_{F}=0.54<1 \). 
We now use the Born approximation (\ref{t1-scat}) and write \emph{\( \lambda ^{(1)}_{\mathrm{tot}}=\lambda _{\mathrm{dir}}^{(1)}-\lambda ^{(1)}_{\mathrm{ex}} \)}. 
We find for the direct part
\begin{eqnarray}
\lambda ^{(1)}_{\mathrm{dir}}=\int _{0}^{2\pi }d\theta |f(\theta )|^{2} &  & \nonumber \\
 &  & \hspace {-40mm}=\lambda _{F}\frac{\tan \gamma }{\cos \gamma }\left\{ 1-2\sin \gamma \tan \gamma \mathrm{Arcth}\left( \sqrt{\frac{1-\sin \gamma }{1+\sin \gamma }}\right) \right\} \label{1} 
\end{eqnarray}
with the Fermi wavelength \( \lambda _{F}=2\pi /k_{F} \), and we
recall that \( \sin \gamma =r_{s}/\sqrt{2}. \) The exchange term
is\emph{\begin{eqnarray}
\hspace {-5mm}\lambda ^{(1)}_{\mathrm{ex}}=\mathrm{Re}\int _{0}^{\pi }d\theta f(\theta )f(\theta -\pi ) &  & \nonumber \\
 &  & \hspace {-48mm}=\lambda _{F}\frac{\sin ^{2}\gamma }{\cos 2\gamma }\left\{ \log \! \left( \! \sin \! \gamma \! +\! \frac{1}{\sin \! \gamma }\right) \! -\! \tan \! \gamma \mathrm{Arcth}\left( \cos \gamma \right) \right\} \label{1} 
\end{eqnarray}
}which \emph{}yields \( \lambda ^{(1)}_{\mathrm{total}}=11.0\, \, \mathrm{nm}. \)
We see here that the Born approximation significantly overestimates the
exact result (\ref{scat tot ex}). For small \( r_{s} \), we can
further approximate \( \lambda ^{(1)}_{\mathrm{dir}}\simeq \lambda _{F}r_{s}/\sqrt{2} \)
and \emph{\( \lambda ^{(1)}_{\mathrm{ex}}\simeq (\lambda _{F}r_{s}^{2}/2)\log (\sqrt{2}/r_{s}) \)},
which gives an even greater length, \( \lambda _{\mathrm{tot}}^{(1)}\simeq 17.2\, \, \mathrm{nm} \).
Note the divergence in the \( r_{s}\to 0 \) limit due to the forward
scattering (\( q=0 \)) divergence of the unscreened Coulomb potential.

\subsection{Angular dependence}

\begin{figure}[t]
{\centering \resizebox*{0.95\columnwidth}{!}{\includegraphics{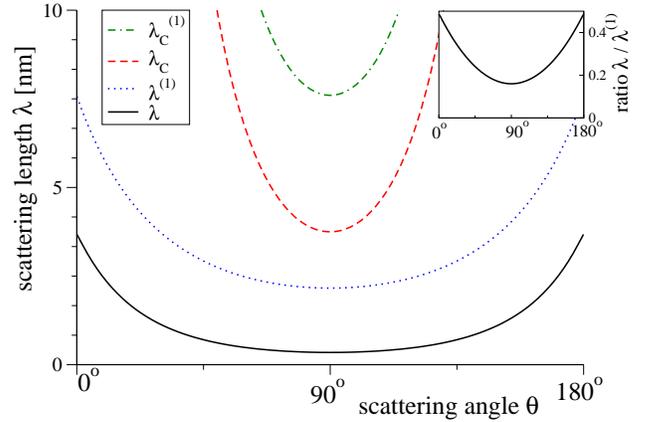}} \par}

\caption{\label{scat-thet}Scattering length \protect\( \lambda (\theta )\protect \)
for a GaAs 2DEG with sheet density \protect\( n=4\cdot 10^{15}\, \mathrm{m}^{-2}\protect \)
and \protect\( k_{B}T=E_{F}/100=2\, \mathrm{K}\protect \), for unpolarized
incident electrons. We compare the exact result (\ref{t-unpol}) for
a 2D Fermi gas to its Born approximation \protect\( \lambda ^{(1)}\protect \),
and to the bare scattering result (with no screening) \protect\( \lambda _{C}\protect \)
and \protect\( \lambda _{C}^{(1)}\protect \); see Eq. (\ref{t1-scat}).
The main effects of the many-body background is to remove the \protect\( \theta =0\protect \)
divergence by screening the Coulomb interaction, and to significantly
reduce the Born approximation via the logarithmic corrections in Eq.
(\ref{tmatrix result}). The inset shows the ratio of the exact scattering
length \protect\( \lambda (\theta )\protect \) to the Born approximation
result \protect\( \lambda ^{(1)}\protect \). }
\end{figure}

We compare in Fig. \ref{scat-thet} the angular dependence of the
different scattering lengths for unpolarized electrons. We first see
that the main effect of the Fermi sea is to reduce significantly the
scattering by one order of magnitude compared to the vacuum case.
The large renormalization is related to the relatively large value
of \( r_{s}=0.86 \) (and the large screening \( k_{s}\sim k_{F} \)),
which strongly reduces the forward scattering divergence of the bare
scattering \( \lambda _{C},\lambda _{C}^{(1)}(\theta \to 0) \). Furthermore,
we notice that the Born approximation \( \lambda ^{(1)} \) is not
valid in the Cooper channel, as higher order terms reduce the scattering
amplitude. The fact that higher terms contribute significantly, despite
the weakness of \( V(q) \), is due to their logarithmic enhancement
by the factor \( \nu  \). 

The angular dependence of the exact scattering length \( \lambda  \)
is similar, but not identical to the Born approximation result \( \lambda ^{(1)} \),
as shown in the inset of Fig. \ref{scat-thet}. Importantly, \( \lambda (\theta ) \)
is a smooth, monotonic (for \( \theta <\pi /2 \)) function, so that
the interference mechanism survives for the production of EPR pairs
at \( \theta =\pi /2 \).

\subsection{\protect\( r_{s}\protect \)--dependence and Born approximation}

\begin{figure}[t]
{\centering \resizebox*{0.9\columnwidth}{!}{\includegraphics{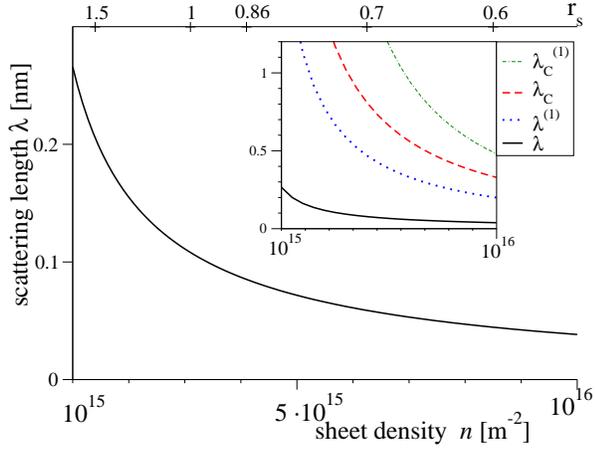}} \par}

\caption{\label{scat-rs}Scattering length \protect\( \lambda \protect \)
at \protect\( \theta =\pi /2\protect \) as a function of the density
\protect\( n\protect \) (see the corresponding \protect\( r_{s}=me^{2}_{0}/\hbar ^{2}\sqrt{\pi n}\protect \)
on the top axis). The inset shows the comparison with the Born approximation
and the bare scattering. }
\end{figure}

We show in Fig. (\ref{scat-rs}) a plot of the scattering length as
a function of the density \( n \) or \( r_{s}=me^{2}_{0}/\hbar ^{2}\sqrt{\pi n} \)
(top axis; we keep \( e_{0} \) constant), for the angle \( \theta =\pi /2 \)
(hence only the singlet channel contributes). There is a strong dependence
\( \sim n^{-1} \) of the scattering, which could be studied experimentally
by varying \( n \) via a top gate. This dependence also roughly applies
to \( \lambda ^{(1)} \) and \( \lambda _{C} \), while \( \lambda _{C}^{(1)}\sim V_{C}^{2}/k_{F}\sim n^{-3/2} \).

The Born approximation \( t(\theta )\simeq v(\theta ) \) is reached
when the logarithmic factor disappears (\( \nu \to 0 \)) and does
not enhance higher order terms; see Eq. (\ref{tmatrix result}). This
occurs at high temperatures \( k_{B}T\to E_{F} \), for hot electrons
\( \xi \simeq E_{F} \), or for electrons that are not in the Cooper
channel (\( P\simeq k_{F} \)). The Born approximation is also reached
in the very small \( r_{s}\sim 0.01 \) limit (not shown), when \( \nu v_{n}\ll 1 \).
For even smaller \( r_{s} \) one can neglect \( k_{s} \) in \( V \),
which yields the Born approximation of the bare Coulomb potential,
\( t\simeq V_{C} \). We note that the effect of the Fermi sea is
intrinsic in our calculation (by restricting the intermediate states
to the Fermi surface), which therefore cannot recover, in the \( r_{s}\to 0 \)
limit, the exact result for the bare Coulomb interaction (\ref{tbare}).

\subsection{Dependence on \protect\( T,\xi \protect \) and \protect\( \alpha \protect \).}

\begin{figure}[t]
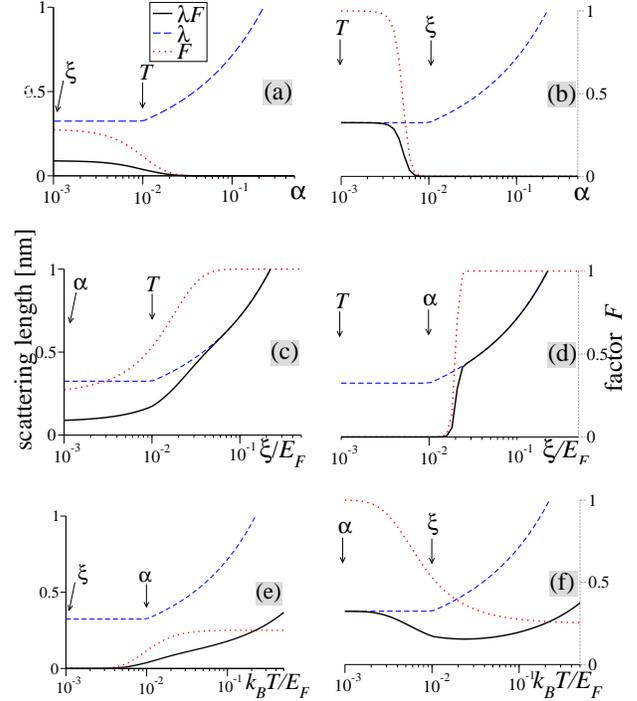

{\centering \resizebox*{0.45\columnwidth}{!}{\includegraphics{Graph/a-x-t.eps}} \( \hspace {2mm} \)\resizebox*{0.45\columnwidth}{!}{\includegraphics{Graph/a-t-x.eps}} \( \vspace {3mm} \)\par}

{\centering \resizebox*{0.45\columnwidth}{!}{\includegraphics{Graph/x-a-t.eps}} \( \hspace {2mm} \)\resizebox*{0.45\columnwidth}{!}{\includegraphics{Graph/x-t-a.eps}} \( \vspace {3mm} \)\par}

{\centering \resizebox*{0.45\columnwidth}{!}{\includegraphics{Graph/t-x-a.eps}} \( \hspace {2mm} \)\resizebox*{0.45\columnwidth}{!}{\includegraphics{Graph/t-a-x.eps}} \par}

\caption{\label{temper}''Observational'' scattering length \protect\( \lambda F\protect \)
at \protect\( \theta =\pi /2\protect \) as a function of the impact
angle \protect\( \alpha \protect \) (a,b), the excitation energy
\protect\( \xi \protect \) (c,d), and temperature \protect\( T\protect \)
(e,f) (left axis). The Fermi occupation factor \protect\( F=[1-n(\xi _{1}')][1-n(\xi _{2}')]\protect \)
(right axis) enforces the requirement that the outgoing states \protect\( \xi _{1,2}'=\xi \pm E_{F}\sin (2\alpha )\protect \)
are available for the outgoing scattering states. The dependence of
the {}``bare'' \protect\( \lambda \protect \) comes from \protect\( \nu \protect \)
(\ref{numax}) and is the same in all graphs. The arrows indicate
the position of the fixed values, e.g. \protect\( \xi /E_{F}=10^{-3},k_{B}T/E_{F}=10^{-2}\protect \)
(a), \protect\( \xi /E_{F}=10^{-2},k_{B}T/E_{F}=10^{-3}\protect \)
(b), etc. }
\end{figure}

Two effects appear when one varies the temperature \( T \), the excitation
energy \( \xi  \) of the incident electrons or the impact
angle \( \alpha =\angle (\mathbf{p}_{1},-\mathbf{p}_{2})/2 \) (a
finite \( |p_{1}-p_{2}| \) plays the same role). The first one is
a change in the factor \( \nu  \) appearing in the denominator \( 1-\nu v_{n} \)
of the \( t- \)matrix (\ref{tmatrix result}). For finite \( \alpha  \),
one should in principle integrate \( \nu (\phi )=(m/2\pi \hbar ^{2})\log \left( 2\alpha \sin |\phi |\right)  \)
over the intermediate angle \( \phi =\angle (\mathbf{k},\mathbf{p}) \)
in the Bethe-Salpeter Eq. \( t(\phi )=v(\phi )+(1/2\pi )\int d\phi \nu (\phi )v(\phi )t(\phi ) \).
However, the dependence of \( \nu (\phi ) \) is smooth (logarithmic)
compared to the behavior of the Coulomb potential at the Fermi surface,
\( v(\phi )\sim 1/(\phi +r_{s}\sqrt{2}) \). Therefore, we neglect
this dependence and set e.g. \( \phi \simeq \pi /3 \) in \( \nu (\phi ) \),
which gives a constant \( \nu \simeq (m/2\pi \hbar ^{2})\log \left( \alpha \right)  \).
Hence we take\begin{equation}
\label{numax}
\nu =\frac{m}{2\pi \hbar ^{2}}\log \left( \mathrm{max}\left\{ \frac{\xi }{E_{F}},\frac{k_{B}T}{E_{F}},\alpha \right\} \right) .
\end{equation}
 The effect of this dependence on the scattering length \( \lambda  \)
at \( \theta =\pi /2 \) is shown in Fig. \ref{temper}. In panel
(a) we fix \( \xi /E_{F}=10^{-3},k_{B}T/E_{F}=10^{-2} \) and vary
\( \alpha  \) (we recall that the Fermi temperature is \( E_{F}/k_{B}=162\, \mathrm{K} \)).
First \( \lambda  \) is constant when \( \alpha <k_{B}T/E_{F} \),
and then increases slowly (the horizontal scale is logarithmic) as
\( \nu  \) decreases; for \( \alpha \to 1 \) \( \then \nu \to 0 \)
we recover the Born approximation \( \lambda ^{(1)} \). The function
\( \lambda  \) is exactly the same if one interchanges \( \xi  \)
and \( k_{B}T \) (b), or permute all the three parameters \( \xi  \),
\( k_{B}T \) and \( \alpha  \) (c-f). 

Secondly, we take into account the requirement that the outgoing states
\( \mathbf{p}_{1,2}' \) are not occupied (and hence available for
the outgoing scattering states), by introducing the factor \( F=[1-n(\xi _{1}')][1-n(\xi _{2}')]=n(-\xi _{1}')n(-\xi _{2}') \),
plotted in Fig. \ref{temper} (right vertical axis). At \( \theta =\pi /2 \)
we have \( p'_{1,2}=p_{1}(\cos \alpha \pm \sin \alpha )\then  \)\( \xi _{1,2}'=\xi \pm E_{F}\sin (2\alpha ) \);
see Fig \ref{setup}(c); hence for a large \( \alpha  \) the final
state \( \xi _{2}'<0 \) will be already occupied and the scattering
into this channel will be prohibited. The transition across the Fermi
surface always occurs at the largest quantity {[}e.g., at \( k_{B}T\gg \xi  \)
in panel (a){]}, and is sharp when temperature is negligible (b,d).
Note that here we consider that the initial states \( \mathbf{p}_{1,2} \)
are always filled (i.e., with occupation \( 1 \)), being either injected
from the QPC or thermally excited.

In (a) the maximum value of \( F \) (when \( \alpha E_{F}\ll k_{B}T \))
is \( F=1/4 \), because the final energies \( |\xi _{1,2}'|\ll k_{B}T \)
lie within the temperature window: \( n(0)=1/2 \). The same occurs
in (c) for \( \xi \ll k_{B}T \), while \( F=1 \) for \( \xi \gg k_{B}T \).
For negligible temperature (b,d) \( F \) saturates to \( 1 \) when
\( \alpha E_{F}\ll \xi  \). We finally note that (for the factor
\( F \)) the panel (e) corresponds to the opposite of panel (a),
(f) is opposite of (c), and (b) is opposite of (d).

We can now consider the combined effect of \( \lambda  \) and \( F \)
by defining the {}``observational'' scattering length \( \lambda F \),
giving the scattering length as could be measured in a real experiment.
As a function of \( \alpha  \), it only reproduces \( F \) by showing
a smooth (a) and sharp (b) step. In (c), it first increases slowly
because of the smooth transition from \( F=1/4\to 1 \) in the region
of constant \( \lambda  \) (small \( \xi  \)), before following
the logarithmic increase of \( \lambda  \) (large \( \xi  \)). In
(d) the transition is sharp and starts at \( \lambda F=0 \). In (e)
the transition is smooth, and the measurable length \( \lambda F \)
follows \( \lambda  \) but is reduced by a factor of \( 4 \). In
(f) there is an interesting non-monotonic behavior in the region
above \( k_{B}T/E_{F}>\alpha  \); however it requires an extremely
small \( \alpha  \), e.g. \( \alpha \simeq 0.2^{\circ } \), not
reachable in a realistic experiment. We note that the rightmost parts
of the graphs (above \( 10^{-1} \)) are only indicative, because
they do not correspond to regime assumed in the derivation of \( \lambda  \)
(\( k_{B}T/E_{F},\xi /E_{F},\alpha \ll 1 \)). 

The scattering length vanishes logarithmically, \( \lambda \sim 1/\log (\nu V)\to 0 \)
when all \( k_{B}T,\xi  \) and \( \alpha \to 0 \). It is reminiscent
of the vanishing of the inverse lifetime of a single quasi-particle
excitation scattering with other electrons below the Fermi surface,
when its excitation energy vanishes \cite{Giu82}. However, the two
cases are completely different: the lifetime diverges because of phase-space
constraints due to the Fermi statistics; in our case the scattering
of two particles \emph{above} the Fermi sea vanishes because of the
renormalization due to the Fermi sea.

\subsection{Quantum oscillations \label{quosc}}

\begin{figure}[t]
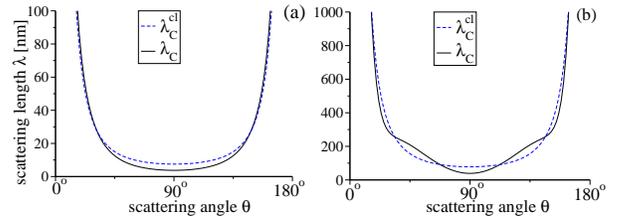

{\centering \resizebox*{0.45\columnwidth}{!}{\includegraphics{Graph/phase-1-1.eps}} \resizebox*{0.45\columnwidth}{!}{\includegraphics{Graph/phase-2-1.eps}} \par}

\caption{\label{phase}Scattering length for the bare Coulomb interaction
{[}given by Eq . (\ref{tbare}){]}, compared to the classical value
given by \protect\( \lambda _{C}^{\mathrm{cl}}=|f_{C}(\theta )|^{2}+|f_{C}(\theta -\pi )|^{2}\protect \).
(a) \protect\( \varsigma =0.6\protect \) corresponding to \protect\( n=4\cdot 10^{15}\, \mathrm{m}^{-2}\protect \).
(b) \protect\( \varsigma =2\protect \) (\protect\( n=4\cdot 10^{14}\, \mathrm{m}^{-2}\protect \)).}
\end{figure}

In addition to the destructive and constructive interference at \( \theta =\pi /2 \),
quantum oscillations can be seen in the \emph{bare} scattering (of
singlets, triplets or unpolarized sources) as a consequence of the
angle--dependent phase \( \varsigma \ln |\sin \theta /2| \) appearing
in Eq. (\ref{tbare}). The number of oscillations is roughly given
by \( \varsigma =r_{s}/\sqrt{2} \), as illustrated in Fig. \ref{phase}.
We see that the oscillations are absent for \( n=4\cdot 10^{15}\, \mathrm{m}^{-2} \),
and only appear at lower density. At \( \theta =\pi /2 \) the quantum
amplitude for unpolarized sources is half of the classical one given
by \( \lambda _{C}^{\mathrm{cl}}=|f_{C}(\theta )|^{2}+|f_{C}(\theta -\pi )|^{2} \),
because the triplet contribution vanishes. 

For the many-particle result, such quantum oscillations could arise
from the small imaginary part appearing with the logarithm in \( \nu (T)=\frac{m}{2\pi \hbar ^{2}}\left( \ln \frac{k_{B}T}{E_{F}}+\frac{\pi }{2}i\right)  \),
as it yields an angle-dependent phase when summing up the Fourier
series. However, the phase is of order \( 1/\ln (k_{B}T/E_{F})\ll 1 \),
and the oscillations are not visible.

\subsection{Production of EPR pairs.}

\begin{figure}[t]
{\centering \resizebox*{0.9\columnwidth}{!}{\includegraphics{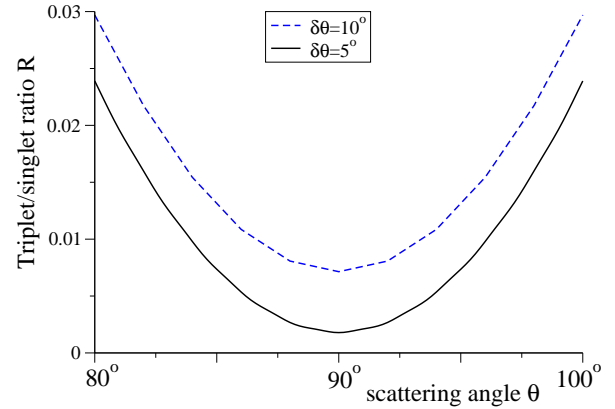}} \par}

\caption{\label{ratio}Ratio \protect\( R(\theta \, \delta \theta )\protect \)
of the number of triplets/singlets collected in the detectors, placed
at an angle \protect\( \theta \protect \) and with an aperture \protect\( \delta \theta =5,10^{\circ }.\protect \)}
\end{figure}

We now consider the setup of Fig. \ref{setup}(b), with detectors
placed at an angle \( \theta \simeq \pi /2 \). The triplet channel
is non-zero because of the small aperture angle \( 2\delta \theta  \)
of the detectors. The scattering lengths for the singlet/triplet channels
into the detectors read\begin{equation}
\label{sintripdet}
\bar{\lambda }_{S/T}(\theta ,\delta \theta )=2\int _{\theta -\delta \theta }^{\theta }d\theta '|f(\theta ')\pm f(\pi -\theta ')|^{2},
\end{equation}
which we use to define the ratio\begin{equation}
\label{rattt}
R(\theta ,\delta \theta )=\frac{N_{T}}{N_{S}}=\frac{3\bar{\lambda }_{T}}{\bar{\lambda }_{S}}
\end{equation}
between the number \( N_{T/S} \) of singlet/triplets collected in
the detectors. 
Here we have allowed for the case where the average scattering angle $\theta$ deviates from  $\pi/2$. 
We show a plot of \( R(\theta ,\delta \theta ) \)
in Fig. \ref{ratio} for \( \delta \theta =5^{\circ } \) and \( 10^{\circ } \).
We find very low values, \( R(90^{\circ },5^{\circ })=0.183\% \),
or 
\( R(85^{\circ },5^{\circ })\simeq R(90^{\circ },10^{\circ })\simeq 0.7\% \),
which shows that the collision entangler is efficient as singlets
are predominantly collected in the detectors, even for \( \theta =80^{\circ } \).
We note that such deviation from \( \pi /2 \) would occur experimentally,
because the electrons injected through the QPCs have an angular spread.
This spread could, however, be reduced by the use of a lense-shaped
top gate implementing a refractive medium for the electron motion
\cite{Roy02}. 

Expanding in \( \delta \theta  \) we find \begin{eqnarray}
\bar{\lambda }_{S/T}(\theta ,\delta \theta )\simeq 2\delta \theta (f\pm f_{\mathrm{ex}})^{2} &  & \nonumber \\
 & \hspace {-50mm}+\frac{2}{3}\delta \theta ^{3}\left[ (f'\mp f'_{\mathrm{ex}})^{2}-(f\pm f_{\mathrm{ex}})(f''\pm f_{\mathrm{ex}}'')\right] +\mathcal{O}(\delta \theta ^{4})\label{1} 
\end{eqnarray}
with\begin{equation}
\label{1}
f\equiv f(\theta )\, \, ,\, \, f_{\mathrm{ex}}\equiv f(\theta -\pi )
\end{equation}
 and \( ' \) denotes \( d/d\theta  \). For \( \theta =90^{\circ },\delta \theta =5^{\circ } \),
and neglecting \( f'' \) we get \begin{equation}
\label{i}
R(90^{\circ },\delta \theta )=\left| \frac{f'(\pi /2)}{f(\pi /2)}\right| ^{2}\delta \theta ^{2}=0.178\%
\end{equation}
which is close to the exact value \( R=0.183\% \) found above. Thus
the Taylor approximation is accurate, and it is clear that the triplet
contribution can be made arbitrarily small by reducing the aperture
\( \delta \theta  \). We note that our calculation gives a ratio
\( |f'/f| \) of the order unity for a wide range of parameter, \( k_{B}T/E_{F}=10^{-1}-10^{-10} \)
and \( r_{s}=0.1-1 \). Using the Born approximation we find a significantly
lower value (\( \delta \theta =5^{\circ } \))

\begin{equation}
\label{ratio-born}
R(90^{\circ },\delta \theta )\simeq \frac{1}{4\left( r_{s}+1\right) ^{2}}\delta \theta ^{2}=0.05\%
\end{equation}
that would be more advantageous for EPR production (see the discussion
on hot electrons in Sec. \ref{hot}).

\subsection{Current}

We now estimate the singlet current collected in the detectors for
a given input current \( I \). We neglect the angular dispersion
of the incident electrons (due to diffraction on the edge of the QPC),
by assuming that the electrons occupy the lowest transverse mode in
the QPC, and that the remaining spread could, in principle, be compensated
by the use of lensing effect \cite{Roy02}. This gives longitudinal
plane waves (with wave vectors \( \mathbf{p}_{1,2} \)) having a transverse
width \( w \) roughly given by half the width of the QPC. 

We first note that the scattering length for the singlet channel is
small, \( \bar{\lambda }_{S}=0.24\, \, \mathrm{nm} \). Taking \( w=100\, \, \mathrm{nm} \),
we find the probability \begin{equation}
\label{4}
P_{S}=\frac{1}{4}\frac{\bar{\lambda }_{S}}{w}=0.06\%
\end{equation}
 for the singlets to be scattered into the detectors. First we assume
that the electrons are injected simultaneously from the reservoirs
(which can be achieved by opening and closing both QPCs at the same
time); this yields a singlet current of \begin{equation}
\label{3}
I_{S}=P_{S}I=0.6\, \, \mathrm{pA}.
\end{equation}
 We have considered a given current of \( I=1\, \, \mathrm{nA} \),
which corresponds to a frequency in the GHz range for the opening
and closing of the QPCs. Otherwise, the electrons are injected at
random times, given by a Poisson process with rate \( W_{\mathrm{in}}=e/I \).
Then the probability of finding two electrons inside the scattering
region (i.e., in state \( \ket {\mathbf{p}_{1},\mathbf{p}_{2}} \))
is roughly \( P_{12}=(W_{\mathrm{out}}/W_{\mathrm{in}})^{2} \), where
\( W_{\mathrm{out}} \) is the rate of escape from the scattering
region into the drain contacts (Fig. \ref{setup}). Finally, we find
the \emph{total} scattering probability of two electrons \begin{equation}
\label{3}
P_{\mathrm{tot}}=\frac{\lambda _{\mathrm{tot}}}{w}=3.4\%
\end{equation}
 (i.e., not necessarily into the detector).

An additional interesting topic is the noise \cite{Naz03} of the
detected current. Since this is outside the scope of the present work,
we only give here heuristic arguments. As the scattering probability
is very small, one can assume that subsequent pairs do not interact
with each other. This implies that the zero-frequency noise induced
by the scattering should be mainly given by the partition noise\[
S(\omega =0)\propto IP_{S}(1-P_{S})\simeq IP_{S},\]
which becomes Poissonian for \( P_{S}\ll 1 \). The pulsed injection
of the electron via e.g. the periodic lowering of the QPC barriers
reduces the stochastic nature of the tunneling through the QPC if
the lowering is sufficiently fast. On the other hand, this periodic
change should lead to a more complex noise behavior for finite frequencies
\cite{Les94}.

\subsection{Hot electrons\label{hot}}

It is interesting to consider the case of hot electrons with larger
excitation energies \( \xi  \) (e.g., a fraction of the Fermi energy
\( E_{F} \) of the scattering region), obtained by applying a dc
bias voltage \( \Delta V \) across the input QPCs. It can be problematic
to have incident electrons with such a wide range of energy, as this
allows a mismatch of the incident energies (\( \xi _{1}\neq \xi _{2} \))
after averaging over both incident energy ranges, which introduces
uncertainties in the scattering angle (see Fig \ref{setup}). To avoid
this situation, one can raise the QPC heights such as to allow only
a very small range of electrons to go above the QPC barrier \cite{Roy02}.
For hot electrons with \( \xi \simeq E_{F} \), the exact result moves
toward the Born approximation; see Eq. (\ref{ratio-born}) and Fig.
\ref{scat-thet}. Hence the scattering length increases {[}because
the logarithmic factor decreases \( \nu \sim \log (\xi /E_{F}) \){]},
while the triplet/singlet ratio becomes more favorable (i.e. smaller).
On the other hand, the scattering length becomes smaller for higher
momentum {[}as \( V\sim 1/(k+k_{s}) \){]}. Taking e.g. \( e\Delta V=3\, \mathrm{meV}\simeq E_{F}/5 \)
\cite{Roy02}, one finds values that are more favorable than for
cold electrons: the singlet length is doubled \( \bar{\lambda }_{S}^{(1)}=0.56\, \mathrm{nm} \),
while the triplet/singlet ratio is halved \( \mathcal{R}=0.10\% \).
Note that hotter electrons have a smaller lifetime because of the
increased phase-space that is available for scattering with electrons
below the Fermi surface \cite{Giu82}. Estimates of the electron-electron
scattering length \( l_{e-e} \) have been obtained \cite{Roy02}
for a GaAs 2DEG using imaging techniques via an SPM, in good agreement
with theoretical predictions \cite{Giu82}. In our case, one has \( l_{e-e}\simeq 1.2\, \mu \mathrm{m} \),
which is similar to the scale \( L\simeq 1\, \mu \mathrm{m} \) of
our envisioned setup. Hence, one can expect some reduction of the
signal due to relaxation into the Fermi sea, roughly given by \( \sim e^{-L/l_{e-e}}\sim 0.3 \).

\subsection{Detection of entanglement}

An important question is to demonstrate that the collected electrons
are indeed spin-entangled EPR pairs. We propose here three ways to
answer this question experimentally. The first one is to refocus the
scattered electrons into a beam splitter and carry out noise measurements
in one outgoing lead; in this situation 
enhanced noise (bunching) is a signature of
the desired singlet state, while zero noise corresponds to entangled
or unentangled triplets \cite{Bur03}. However, this method would
probably require some bridges to avoid the source reservoirs. The
second one is to carry out tests of violation of Bell inequality \cite{Kaw01,Gran04,Sam03},
by measuring single-spin projections via a single electron transistor
coupled to a spin filtering device. The latter can be a quantum dot
in the Coulomb blockade regime\cite{Han03} or a QPC \cite{Pot02}
in a strong in-plane magnetic field. The third method consists in
adding a p-i-n junction \cite{Fie99,Nor99}, allowing the recombination
of the entangled electrons with unentangled holes into photons; one
should then carry out the test of Bell inequality with the photons,
by measuring their entangled polarization modes. 

In addition, we mention that the interference mechanism responsible
for the vanishing of the triplets at \( \pi /2 \) could be demonstrated
by polarizing the incoming electrons spin, which can be achieved by
applying a large in-plane magnetic field to turn the QPCs into spin
filters \cite{Pot02}, or by replacing them by quantum dots \cite{Han03}.
The current recorded at the detectors, which is proportional to the
fraction of incoming singlets \( \rho _{S}=(1-\mathcal{P}^{2})/2 \),
 should then rapidly decrease as the polarization \( \mathcal{P} \)
of the spins increase.

\subsection{Creation of localized, non-mobile entanglement}

\begin{figure}[t]
{\centering \resizebox*{0.7\columnwidth}{!}{\includegraphics{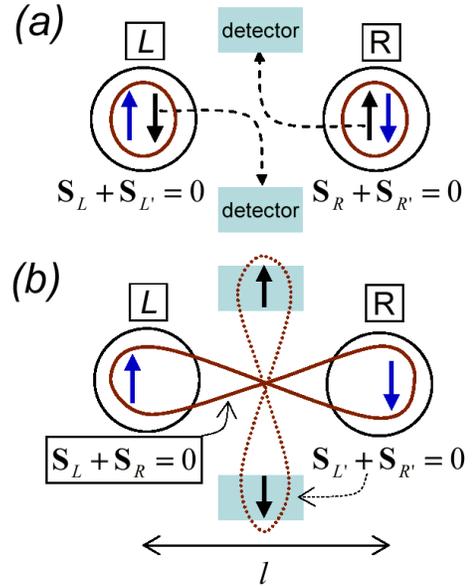}} \par}
\caption{\label{nonmobile}Creation of non-mobile entanglement. Each input
QPC is replaced by a quantum dot (\protect\( L\protect \) and \protect\( R\protect \))
containing two excess electrons. The ground state in each dot is the
singlet with total spin \protect\( \mathbf{S}_{L}+\mathbf{S}_{L'}=\mathbf{S}_{R}+\mathbf{S}_{R'}=0\protect \)
(a). One injects one electron from each dot (e.g. \protect\( L'\protect \)
and \protect\( R'\protect \)), and allows them to scatter. When
they are detected at a \protect\( \pi /2\protect \) scattering angle,
we know that they are in the singlet state \protect\( \mathbf{S}_{L'}+\mathbf{S}_{R'}=0\protect \).
(b). As the total spin is conserved, the two remaining electrons in
dots \protect\( L\protect \) and \protect\( R\protect \) are also
in the singlet state \protect\( \mathbf{S}_{L}+\mathbf{S}_{R}=0\protect \),
and therefore form a localized EPR pair, whose members are separated
by the interdot distance \protect\( l\protect \).}
\end{figure}

We discuss here a way to produce static spin-entangled electrons,
described in Fig \ref{nonmobile}. We propose to replace the two {}``injection''
QPCs by two quantum dots, each with an even number of electrons, so
that two excess electrons are in the singlet ground state \cite{Tar00}.
Lowering the tunneling barrier defining the dots allows for the simultaneous
injection of one electron (of the singlet pair in each dot) into the
scattering region. If these electrons are detected at \( \pi /2 \)
by the detector, we know that they must be in a spin-singlet state
(with certainty \( 1-N_{T}/N_{S} \)). As spin is conserved during
the Coulomb scattering, the total spin for the two electrons left
in the two dots must be zero, which corresponds to the spin singlet
state. Therefore, one has created a localized (non-mobile) entangled
pair of electrons separated by an interdot distance \( l\sim 1\, \mu \mathrm{m} \).
This instantaneous {}``creation'' (obtained by post-selection) is
a dramatic illustration of Einstein's {}``spooky action at a distance''.
Contrary to the standard EPR paradox, it does not manifest itself
in the results of measurement correlations, but in the {}``creation''
of a non-local (in the sense of non-overlapping wavefunctions) quantum
state. Such process, similar in some sense to entanglement swapping or
quantum teleportation,
could be useful in a scalable quantum computer to create entangled
pairs without having to go through the standard sequence of swapping
state, which requires to move one electron to neighboring site of
the other one, entangle them by local interaction \cite{Los98} and
move the electron back to its original place. However, we note that
our scheme requires a large number of collisions, of the order of
\( 1/P_{S} \); this number scales fortunately more slowly
(\( \sim 1/\delta \theta  \))
with the aperture \( \delta \theta  \) than the precision (\( \sim N_{S}/N_{T}\sim 1/\delta \theta ^{2} \)).

\section{Electron-phonon and electron-electron interaction\label{sec phon}}

In this section we investigate the question whether phonons can influence
the scattering amplitude in a significant way. We first note that
the scattering of electrons on real phonons can be neglected here,
as it is strongly suppressed at low temperature. This is illustrated
for instance by the absence of phonon effects in the experiments of
Ref.\cite{Top00,Roy02}. However, the effective electron-electron
interaction arising from the exchange of virtual phonons does not
depend on temperature, so that it could play a role in the electron-electron
scattering. Our goal here is to estimate it and compare it to 
the screened Coulomb
interaction which we have considered so far. We shall see that the
contribution of acoustic phonons (deformation and piezoelectric coupling)
is negligible, while the polar phonons give a smooth monotonic decrease
of the electron-electron interaction that is less than \( 20\% \),
and as such does not change qualitatively the results presented in
Sec. \ref{sec res}.

\subsection{2D phonon-mediated electron-electron interaction}

In 3D, the effective electron-electron interaction is given by \cite{Mah00}
\begin{equation}
\label{3}
H^{\mathrm{ph}}_{\mathrm{e}-\mathrm{e}}=\frac{1}{2V}\sum _{\vec{k},\vec{k}'}\sum _{\vec{q}}c_{\vec{k}+\vec{q}}^{\dagger }c_{\vec{k}'-\vec{q}}^{\dagger }c_{\vec{k}'}c_{\vec{k}}W^{\mathrm{ph}}(\vec{q}),
\end{equation}
 where \( V \) the normalization volume, \( \vec{q}=(\mathbf{q},q_{z}),\vec{k},\vec{k}' \)
are 3D vectors and \( \mathbf{q},\mathbf{k},\mathbf{k'}\) are 2D vectors
in the plane of the 2DEG (in the following we keep the notation \( q=|\mathbf{q}| \)).
The electron-electron interaction matrix element reads \begin{equation}
\label{ephe}
W^{\mathrm{ph}}(\vec{q})=|M(\vec{q})|^{2}\frac{2}{\hbar V}\frac{\omega _{\mathrm{ph}}(q)}{\omega ^{2}-\omega ^{2}(q)},
\end{equation}
where \( \omega _{\mathrm{ph}}(q) \) is the phonon dispersion, and
\( M(\vec{q}) \) is the matrix element for the electron-phonon interaction\begin{equation}
\label{21}
H_{\mathrm{e}-\mathrm{ph}}=\frac{1}{V}\sum _{\vec{q}}\sum _{\vec{k}\sigma }c_{\vec{k}+\vec{q},\sigma }^{\dagger }c_{\vec{k}\sigma }(b_{\vec{q}}+b_{-\vec{q}}^{\dagger })M(\vec{q}).
\end{equation}
Here \( b^{\dagger }_{\vec{q}} \) and \( c^{\dagger }_{\vec{k}} \)
are phonon and electron creation operators. We shall consider the
lowest order in \( W^{\mathrm{ph}} \), which in the Cooper channel
allows us to take the static limit \( \omega =0 \) as all the energies
involved in the scattering are the same, \( E_{i}\simeq E_{F} \). 

The electron-phonon interaction \( M(\vec{q}) \) as well as the effective
interaction \( W^{\mathrm{ph}}(\vec{q}) \) are always 3D as they
involves coupling of the 2D electrons with the bulk 3D phonons\emph{.}
There is no 3D screening of the bare ion-electron Coulomb interaction,
as there are no mobile charges in the bulk. Now we define an effective
2D interaction \( W_{2D}(\mathbf{q}) \), which we shall compare to
the unscreened 2D Coulomb interaction \( V_{C} \). We assume that
the electron wave function is separable into a plane wave \( \ket {\mathbf{k}} \)
and a confined lateral function \( \ket {\psi } \). For instance,
one can take an infinite square well of width \( L \) \begin{equation}
\label{l}
\psi (z)=\frac{2}{L}\sin \left( \frac{\pi z}{L}\right) 
\end{equation}
which yields the width of the 2DEG \begin{equation}
\label{l}
d=\left\langle z^{2}-<z^{2}>\right\rangle ^{1/2}=L\sqrt{\frac{1}{2\pi ^{2}}-\frac{1}{6}}\simeq 0.18L.
\end{equation}
We prefer to consider the alternative variational solution of the
triangular well present at the interface \cite{And82}, \begin{equation}
\label{triangwf}
|\psi (z)|^{2}=\frac{1}{2}\kappa ^{3}z^{2}e^{-\kappa z}
\end{equation}
with the width\begin{equation}
\label{ll}
d=\frac{\sqrt{3}}{\kappa },
\end{equation}
as it allows for simple analytical expressions. We define the effective
\( W^{ph}_{2D}(\mathbf{q}=\mathbf{k}-\mathbf{k}') \) by \begin{equation}
\label{l}
\bra {\mathbf{k}_{1}',\psi ;\mathbf{k}_{2}',\psi }W^{\mathrm{ph}}\ket {\mathbf{k}_{1},\psi ;\mathbf{k}_{2},\psi }=\delta (\mathbf{k}_{1}+\mathbf{k}_{2}-\mathbf{k}_{1}'-\mathbf{k}_{2}')W^{\mathrm{ph}}_{2D}(\mathbf{q})
\end{equation}
and get \cite{Bon98}\begin{eqnarray}
W^{\mathrm{ph}}_{2D}(\mathbf{q})=\int dz\int dz'|\psi (z)|^{2}|\psi (z')|^{2}W^{\mathrm{ph}}(\mathbf{q};z-z') &  & \nonumber \\
 &  & \hspace {-60mm}=\frac{1}{2\pi }\int dq_{z}W^{\mathrm{ph}}(\mathbf{q},q_{z})\left| I(q_{z})\right| ^{2},\label{d2d3} 
\end{eqnarray}
with \( W^{\mathrm{ph}}(\mathbf{q},z)=(1/2\pi )\int dqW^{\mathrm{ph}}(\mathbf{q},q_{z})e^{-iqz} \)
and the form factor \begin{equation}
\label{l}
I(q_{z})=\int dze^{iq_{z}z}|\psi (z)|^{2}.
\end{equation}
 The latter is particularly simple for the triangular well, \( I(q_{z})=\left( iq_{z}/\kappa -1\right) ^{-3} \).
Our goal is to find the strength of this additional e-e interaction
relative to the unscreened Coulomb potential \( V_{C} \), by defining
the ratio\begin{equation}
\label{ratioph}
r=\frac{W^{\mathrm{ph}}_{2D}}{V_{C}}(q=k_{F}).
\end{equation}

\paragraph*{Parameters for GaAs}

We consider a well of width \( d=5\, \, \mathrm{nm} \), and take
the following parameters \cite{ioffe} for GaAs: the mass density
\( \rho _{m}=5320 \) kg/m\( ^{3} \), the deformation potential constant
\( D=-7\, \mathrm{eV} \), the piezoelectric constant \( eh_{14}=1.44\cdot 10^{9}\, \mathrm{eV}/\mathrm{m} \),
the acoustic sound velocity \( c=3700 \) m/s (we assume here that
\( c \) is the same for both longitudinal and transverse phonons),
the optical (longitudinal and transverse) phonon frequencies \( \hbar \omega _{LO}=36.6\, \, \mathrm{meV} \),
\( \hbar \omega _{TO}=33.8\, \, \mathrm{meV} \), the ionic plasmon
frequency: \( \Omega _{p,i}=85.5\, \, \mathrm{meV} \) and, finally,
the low- and high-frequency dielectric constants \( \epsilon (0)= \)12.9,
\( \epsilon (\infty ) \) =10.89.

\subsection{Acoustic phonons: coupling to the deformation potential}

We first consider electrons coupled to the acoustic phonons via the
deformation potential. The electron-phonon matrix element is \cite{Pri81}\begin{equation}
\label{acoudef}
M(\vec{q})=D\sqrt{V\frac{\hbar }{2\rho _{i}c}|\vec{q}|}.
\end{equation}
where \( D \) is the deformation constant, \( \rho _{i} \) is the
mass density, and the dispersion relation is \( \omega _{\mathrm{ph}}(\vec{q})=c|\vec{q}| \).
The static effective e-e interaction is a constant\begin{equation}
\label{i}
W^{\mathrm{ph}}(\vec{q})=-\frac{D^{2}}{\rho _{i}c^{2}},
\end{equation}
which yields in 2D for the triangular well (\ref{triangwf})\begin{equation}
\label{l}
W^{\mathrm{ph}}_{2D}(\mathbf{q})=-\frac{D^{2}}{\rho _{i}c^{2}}\frac{3\kappa }{16}.
\end{equation}
The ratio (\ref{ratioph}) becomes\begin{equation}
\label{l}
r(q)=-q\frac{D^{2}}{\rho _{i}c^{2}e^{2}}\frac{3\sqrt{3}}{16d}\stackrel{q=k_{F}}{=}-1.4\cdot 10^{-3},
\end{equation}
which shows that the effective interaction \( W^{\mathrm{ph}}_{2D} \)
can be neglected for deformation potential coupling.

\subsection{Acoustic phonon: piezoelectric coupling}

For piezoelectric coupling, the matrix element reads \cite{Pri81}\begin{equation}
M(\vec{q})=\frac{eh_{14}}{\epsilon _{r}}\sqrt{V\frac{\hbar }{2\rho _{i}\omega _{\mathrm{ph}}(q)}A(\vec{q})},
\end{equation}
with the \emph{}polarization constant \( eh_{14} \) and the anisotropy
factor \begin{equation}
\label{i}
A(\vec{q})=\left\{ \begin{array}{ccc}
9q_{z}^{2}q^{4}/2|\vec{q}|^{6} &  & (\mathrm{LA})\\
(8q_{z}^{4}q^{2}+q^{6})/4|\vec{q}|^{6} &  & (\mathrm{TA})
\end{array}\right. 
\end{equation}
for longitudinal (LA) or transverse (TA) phonons. It can be replaced
by \( A_{\mathrm{LA}}=0,A_{\mathrm{TA}}=1/4 \) for a 2D system constraining
momentum transfers to \( q_{z}=0 \). This gives\begin{equation}
\label{acoupiezo}
M_{q}=\frac{eh_{14}}{\epsilon _{r}}\sqrt{V\frac{\hbar }{8\rho _{i}c|\vec{q}|}}.
\end{equation}
 The static e-e interaction is therefore proportional to the 3D Coulomb
interaction:\begin{equation}
\label{i}
W^{\mathrm{ph}}(\vec{q})=-\frac{1}{\rho _{i}}\left( \frac{h_{14}}{2c\epsilon _{r}}\right) ^{2}\frac{e^{2}}{q^{2}}.
\end{equation}
 Performing the transformation (\ref{d2d3}), we find for small \( q\alt k_{F}/10 \)
the effective 2D potential:\begin{equation}
\label{i}
W_{2D}^{\mathrm{ph}}(\mathbf{q})=-\frac{1}{\rho _{i}}\left( \frac{h_{14}}{2c\epsilon _{r}}\right) ^{2}\frac{e^{2}}{2q}.
\end{equation}
 This corresponds to the 3D \( \to  \) 2D transformation of the Coulomb
potential, i.e. \( 1/|\vec{q}|^{2}\to 2/q \) . Finally we get\begin{equation}
\label{ll}
r=-\frac{\pi \epsilon _{0}}{\rho _{i}\epsilon _{r}}\left( \frac{h_{14}}{c}\right) ^{2}=-1.5\cdot 10^{-5}.
\end{equation}
Hence the 2D piezoelectric contribution is also negligible.

\subsection{Optical phonons: polar coupling }

\begin{figure}[t]
{\centering \resizebox*{0.9\columnwidth}{!}{\includegraphics{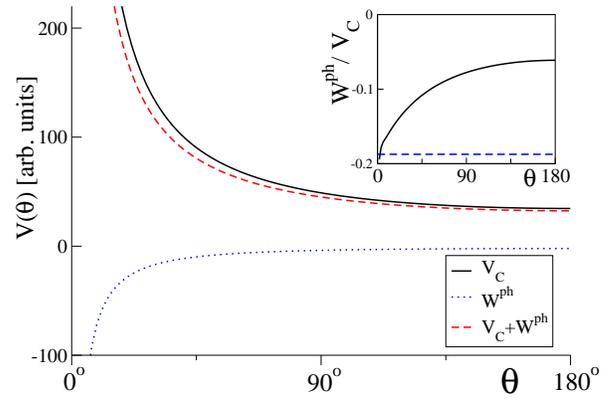}} \par}

\caption{\label{figphon}Effective 2D e-e interaction \protect\( W^{\mathrm{ph}}_{2D}\protect \)
from LO polar phonons as a function of the scattering angle \protect\( \theta \protect \).
We take a well with a width \protect\(d= 5\, \mathrm{nm}\protect \), and fix  \protect\( q=2k_{F}\sin (\theta /2)\protect \) and \protect\( \omega =0\protect \).
We compare it to the unscreened Coulomb interaction \protect\( V_{C}\protect \).
Inset: Ratio \protect\( \mathcal{R}=W_{2D}^{\mathrm{ph}}/V_{C}\protect \)
from (\ref{unsw}) ; the horizontal line corresponds to the small--\protect\( q\protect \)
approximation (\ref{ratunsc}).}
\end{figure}

The electron-phonon matrix element is \cite{Mah00}\begin{equation}
\label{polmatel}
M(\vec{q})=\sqrt{\left[ \frac{1}{\epsilon (\infty )}-\frac{1}{\epsilon (0)}\right] \frac{2\pi e_{0}^{2}}{|\vec{q}|^{2}}\hbar \omega _{LO}V},
\end{equation}
where \( \epsilon (0) \) and \( \epsilon (\infty ) \) are the static
and high-frequency dielectric constants. For \( \omega =0 \), this
yields \begin{equation}
W^{\mathrm{ph}}(\vec{q})=-\left[ \frac{1}{\epsilon (\infty )}-\frac{1}{\epsilon (0)}\right] \frac{4\pi e_{0}^{2}}{|\vec{q}|^{2}}.
\end{equation}
Hence the effective 2D potential is, for small \( q \), \begin{equation}
\label{1}
W^{\mathrm{ph}}_{2D}(q)\simeq -\left[ \frac{1}{\epsilon (\infty )}-\frac{1}{\epsilon (0)}\right] \frac{2\pi e_{0}^{2}}{q},
\end{equation}
and we get the ratio\begin{equation}
\label{ratunsc}
r\simeq -\epsilon _{r}\left[ \frac{1}{\epsilon (\infty )}-\frac{1}{\epsilon (0)}\right] =-19\%.
\end{equation}
 As the ratio \( |r| \) is{ rather large, it is important to consider
here the more accurate expression, valid for larger \( q \), found
by performing the integration (\ref{d2d3}) with the triangular well
solution (\ref{triangwf}):\begin{eqnarray}
W^{\mathrm{ph}}_{2D}(q)=-\left[ \frac{1}{\epsilon (\infty )}-\frac{1}{\epsilon (0)}\right] \frac{2\pi e_{0}^{2}}{q} &  & \nonumber \label{unsw} \\
 &  & \hspace {-60mm}\times \frac{\kappa }{8\left( \kappa ^{2}-q^{2}\right) ^{3}}\left( -3q^{5}+10q^{3}\kappa ^{2}-15q\kappa ^{4}+8\kappa ^{5}\right) ,\label{optlargeq} 
\end{eqnarray}
 which is plotted in Fig. \ref{figphon}. We find the rather unexpected
result that virtual (optical) phonons give a significant contribution:
the ratio is \( |r|<20\% \). However, the effect of \emph{\( W^{\mathrm{ph}}_{2D} \)}
is monotonic and will not change qualitatively the scattering of two
electrons in a GaAs 2DEG.

\section{Kohn-Luttinger instability\label{kohnlut}}

Having found the scattering vertex in lowest order, we now consider
higher-order diagrams and examine whether superconducting fluctuations
could have an effect on the scattering. It has been known for a long
time \cite{Lut66} that in 3D the second-order crossed diagram \( \Lambda _{3} \) in the irreducible vertex (see Fig.\ref{diagram}) can lead to a pairing
instability and a transition to superconductivity. The origin lies
in the susceptibility \( \chi ^{0} \) entering \( \Lambda _{2,3} \);
being non-analytic, its spherical harmonics have a polynomial asymptotic
decay \( \chi ^{0}_{l}\sim l^{-4} \) with respect to the coefficient
\( l \) of the spherical harmonics decomposition, while the single
interaction is analytic and therefore yields \( v_{l}\sim e^{-l} \).
As \( \chi _{l}^{0} \) oscillates, 
the irreducible
vertex becomes attractive 
for sufficiently large \( l \): \( \Lambda _{l}\sim v_{l}^{0}+\Lambda _{3,l}<0 \).
The transition temperature is found from the Cooper divergence of
\( \Gamma  \), i.e. by the relation \( 1=\nu _{3D}\Lambda _{l} \),
where \( \nu _{3D} \) is the same as (\ref{nut}) with the 2D density
of state \( m/2\pi \hbar ^{2} \) replaced by the 3D one. This yield
an infinitesimal temperature \cite{Lut66}, \( k_{B}T_{c}\sim \exp (-10^{5}) \)
for a metal with \( r_{s}=4.5 \).

In 2D the equivalent transition does not occur, because there is no
instability for particles below the Fermi surface:
\( q<2k_{F} \) and  \( \chi ^{0}=-N_{0} \) has no (negative) harmonics.
It has been shown, however, that higher-order diagrams (in \( \Lambda  \))
can lead to a transition \cite{Chu93}. Alternatively, finite energy
transfers can induce {}``pseudo-pairing'' in the \emph{d-}wave Cooper
channel \cite{Gal03}. In our work, however, we are not interested
in a superconducting transition; rather, we would like to verify that
the scattering vertex for the injected particles (which are above
the Fermi surface) is not substantially renormalized by the standard
(lowest-order) Kohn-Luttinger instability with no energy transfer. 

The singular part of \( \Lambda _{2,3} \) originates from the function
\( B(\mathbf{k},\tilde{q}) \) evaluated near \( \tilde{q}=0 \).
Neglecting the variations of \( V \) in the \( q \)--integrals (\ref{lam2}-\ref{lam3})
(this corresponds to approximating \( V(r) \) by a very short range
potential, e.g. a \( \delta  \)--function), we evaluate \( V \)
at the singular points of \( B \) and write\begin{equation}
\label{1}
\Lambda _{2}\simeq 2V(0)V(2k_{F})\int d\mathbf{k}_{1}B(\mathbf{k}_{1},\tilde{q})=-V(0)V(2k_{F})\chi ^{0}(\tilde{q})
\end{equation}
\begin{equation}
\label{1}
\Lambda _{3}\simeq V^{2}(0)\int d\mathbf{k}_{1}B(\mathbf{k}_{1},\tilde{Q})=-V^{2}(0)\chi ^{0}(\tilde{Q})/2
\end{equation}
where \( q=|\mathbf{p}'-\mathbf{p}|\simeq 2k_{F} \) and \( Q=|\mathbf{p}'+\mathbf{p}|\simeq 2k_{F} \).
We see that we can neglect \( \Lambda _{2}\sim r_{s}\Lambda _{3} \);
we also take the static limit (\ref{pol2d}) of \( \chi ^{0}(\omega ) \)
as it varies on a scale \( \sim E_{F} \). 
As $Q$ should be slightly above 
$2 k_F$, we take
\begin{equation} \label{1}
Q=k_{F}(2+c)\cos \theta /2
\end{equation}
with \( c= \xi /E_{F} \simeq 2(p-k_F)/k_F \ll 1 \) and \( |\theta |\ll 1 \).
The Fourier coefficients of the crossed diagrams are therefore given
by
\begin{eqnarray}
\Lambda _{3,n}=V^{2}(0)\frac{1}{2\pi }\int _{0}^{\pi }d\theta \cos (n\theta )\frac{m}{\pi \hbar ^{2}} &  & \nonumber \\
 &  & \hspace {-50mm}\times \left[ 1-\Theta (\bar{\theta }-\theta )\sqrt{1-\frac{1}{(1+c/2)^{2}\cos ^{2}(\theta /2)}})\right] \label{1} 
\end{eqnarray}
with \( \bar{\theta }\simeq \sqrt{2c} \). In lowest order in \( c \),
we find for the singular part\begin{equation}
\label{1}
\Lambda _{3,n}=V^{2}(0)\frac{m}{2\pi \hbar ^{2}}\frac{\bar{\theta }}{4n}J_{1}(n\bar{\theta }),
\end{equation}
and the asymptotics\begin{equation}
\label{l}
\Lambda _{3,n}\stackrel{n\gg 1}{\simeq }V^{2}(0)\frac{m}{8\pi \hbar ^{2}}\sqrt{\frac{\pi \bar{\theta }}{n^{3}}}\sin (n\bar{\theta }-\pi /4),
\end{equation}
 compared \cite{Lut66} to \( l^{-4} \) in 3D. Note the oscillatory
behavior, which allows for negative values. The instability temperature
can be estimated by requiring \begin{equation}
\label{ncrit}
|\Lambda _{3,n}|\ge v_{n}\then n>n_{0}=\frac{4}{\pi r_{s}^{2}\bar{\theta }},
\end{equation}
where we neglected the oscillating sine function, and used (\ref{vnasym})
for \( v_{n} \). We find\begin{equation}
\label{l}
k_{B}T\sim E_{F}e^{-1/|\Lambda _{3,n_{0}}|}\simeq E_{F}e^{-4/r_{s}^{3}c}.
\end{equation}
Note that the parameter \( c \) appears in 2D because of the \( \Theta  \)
function in (\ref{pol2d}), and is absent in 3D. For GaAs and taking
\( c=0.02 \), we find \( k_{B}T/EF\sim e^{-100} \), which means
that the attractive effect of the crossed diagram is completely negligible,
and does not lead to any sizable fluctuations of the scattering vertex.
For a metal with \( r_{s}\simeq 4.5 \), the transition temperature
in 3D was found \cite{Lut66} to be \( \sim E_{F}e^{-40000} \); in
2D we cannot neglect the sine as in (\ref{ncrit}); numerically we
find the temperature \( \sim E_{F}e^{-20/c} \). This can be larger
than in 3D for \( c>10^{-4} \), despite the fact that the asymptotic
decay of \( v_{n}\sim n^{-2} \) is much slower than in 3D (\( v_{l}\sim e^{-l} \)).

\section{Conclusion}

The prospect of experiments probing individual electron collisions
in a 2DEG is a natural motivation to study the problem of two electron
interacting via Coulomb interaction in the presence of a Fermi sea.
One of the main result of this work is the expression (\ref{tmatrix result})
for the scattering amplitude for two electrons in the Cooper channel.
We found that the presence of the Fermi sea yields a significant renormalization
of the strength of the scattering, rather similar to the renormalization
found in the discussion of the Cooper instability. This is closely
linked to the selection of intermediate states at the Fermi surface.
Away from the Cooper channel, this selection disappears, and the Born
approximation is valid. The overall angular dependence is fairly unmodified
and smooth, There is a sizable dependence on the sheet density, while
the dependence on temperature, energy and impact angle is strongly
influenced by the Fermi occupation factors. The total scattering length
is \( \lambda _{\mathrm{tot}}\simeq 3\, \mathrm{nm} \), which is
of the same order as the Fermi wavelength.

We discussed how to use such collisions to produce EPR pairs at a
scattering angle of \( \theta =\pi /2 \). This mechanism is rather
robust against imprecisions in \( \theta  \), for an output singlet
current around \( 0.5\, \mathrm{pA} \). The EPR production was found
to be slightly more efficient in the case of hot electrons. We discussed
detection of entanglement and quantum interference, and proposed a
way to create localized EPR pairs separated by mesoscopic distances. 

We studied phonon-mediated electron-electron interaction. We found
that the dominant contribution comes from polar coupling to optical
phonons, but does not affect qualitatively the Coulomb scattering.
The strength of the Kohn-Luttinger superconducting instability was
calculated, and shown to be negligible. Finally, we developed (in
Appendix \ref{sec smallrs}) an alternative calculation valid for
diverging forward scattering contributions, and showed them to be
negligible in GaAs.

\begin{acknowledgments}
We thank C. Egues, V. Golovach, W. Coish, A. Bleszynski, B. Lee and
M. Yildirim for useful discussions. This work has been supported by
NCCR ``Nanoscale Science'', Swiss NSF, EU-Spintronics, DARPA, ARO,
and ONR.
\end{acknowledgments}
\appendix

\section{Small $r_s$-approximation at \protect\( \theta =\pi /2\protect \)\label{smallrsapprox}}

Here we derive analytical expressions for the \emph{t-}matrix and
its derivative at \( \theta =\pi /2 \), which we then use to compute
the ratio \( R \). We first note that \( t(\pi /2) \) is an alternating
series\begin{equation}
\label{1}
t(\pi /2)=t_{0}+2\sum _{n>1}(-1)^{n}t_{2n}
\end{equation}
where \( t_{n}=v_{n}/(1-\nu v_{n}) \). We write the differences as
\( t_{n}-t_{n+2}=h_{n}/[(1-\nu v_{n})(1-\nu v_{n+2})] \) with \( h_{n}=v_{n}-v_{n+2} \),
which allows us to get the smoother series\begin{equation}
\label{1}
t(\pi /2)=\frac{h_{0}}{(1-\nu v_{0})(1-\nu v_{2})}-\frac{h_{2}}{(1-\nu v_{2})(1-\nu v_{4})}+...
\end{equation}
 Now we use for the denominators the very small \( r_{s} \) approximation
(\ref{vnap1}) \( v_{n}\simeq -\bar{v}\log r_{s} \), \( \bar{v}=2e_{0}^{2}/k_{F} \),
giving \begin{equation}
\label{2}
\hspace {-5mm}t(\pi /2)=\frac{1}{(1+\nu \bar{v}\log r_{s})^{2}}\left( h_{0}-h_{2}+h_{4}-h_{6}+...\right) .
\end{equation}
For the numerators we write
\begin{eqnarray}
h_{n} & = & 2\bar{v}\left( \frac{\cos \gamma }{2n+1}+\frac{\cos 3\gamma }{2n+3}\right) \nonumber \\
 &  & \hspace {-10mm}+2\bar{v}\left\{ \frac{\cos 5\gamma -\cos \gamma }{2n+5}+\frac{\cos 7\gamma -\cos 3\gamma }{2n+7}+...\right\} \label{2} 
\end{eqnarray}
and neglect the second term which is order of \( r_{s} \). Then\begin{equation}
\label{2}
t(\pi /2)=\frac{2\bar{v}}{(1+\nu \bar{v}\log r_{s})^{2}}\sum _{n>0}(-1)^{n}\left( \frac{1}{4n+1}+\frac{1}{4n+3}\right) 
\end{equation}
and finally\begin{equation}
\label{tpi2app}
t(\pi /2)=\frac{2\bar{v}}{(1+\nu \bar{v}\log r_{s})^{2}}\frac{\pi }{2\sqrt{2}}.
\end{equation}
This approximation is good for very low \( r_{s} \); the error is
\( \sim 10\% \) for \( r_{s}<0.09 \), which corresponds however
to a very high density \( n=4\cdot 10^{17} \).

We proceed similarly for the derivative \( t'(\pi /2) \):

\begin{equation}
\label{3}
t'(\theta /2)=2\left( -t_{1}+3t_{3}-5t_{5}+7t_{7}-9t_{9}+...\right) 
\end{equation}
\begin{equation}
\label{3}
\simeq \frac{4\bar{v}}{(1+\nu \bar{v}\log r_{s})^{2}}\sum _{n>1}(-1)^{n}n\left( \frac{1}{4n-1}+\frac{1}{4n+1}\right) .
\end{equation}
 The sum yields \( -\pi /8\sqrt{2}+(1/4)(-1)^{N} \) with \( N\to \infty  \).
Neglecting the oscillating term \cite{oscill-negl}, we get\begin{equation}
\label{der-app-2}
t'(\pi /2)=\frac{1}{2}t(\pi /2).
\end{equation}
We find in this approximation a very simple form for the ratio \begin{equation}
\label{5}
R(\theta ,\delta \theta )=\frac{1}{4}\delta \theta ^{2}.
\end{equation}
 This corresponds to the Born approximation result, Eq. (\ref{ratio-born}),
in the limit of no screening (\( r_{s}\to 0 \)). This is somewhat
surprising, as our result (\ref{tpi2app}) still contains both the
screening (finite \( r_{s} \)) and the resumed higher order terms
(responsible for the term \( \nu \bar{v} \) in the numerator). One
must further expand \( t(\pi /2) \) in small \( r_{s} \) in order
to recover the Born approximation with an unscreened potential \begin{equation}
\label{fs-app-4}
t(\pi /2)\simeq \bar{v}\frac{\pi }{\sqrt{2}}=r_{s}\pi \frac{\hbar ^{2}}{m}=V_{C}(q=k_{F}\sqrt{2}).
\end{equation}
The Fourier series is not well defined in this case, because of the
forward-scattering divergence \( V_{C}(0) \). We also note that the
number of Fourier coefficient required to reach convergence of the
numerical Fourier sum increases dramatically to \( n_{max}=70000 \)
for \( n=10^{23} \) (with the heuristic dependence \( n_{max}\sim r_{s}^{-0.8} \)),
as the potential becomes more peaked.

\section{Forward scattering when $r_s\to 0$\label{sec smallrs}}

Here we consider carefully the limit of vanishing \( r_{s}\to 0 \),
by following a different approach to solve the Bethe-Salpeter Eq.
that allow us to study the contribution of forward scattering states.
These are indeed important in the very small \( r_{s} \) limit, as
the \emph{unscreened} Coulomb scattering cross-section has a forward
scattering divergence (i.e., for vanishing momentum transfers, \( q=0 \))
in 3D and 2D.

The calculation which was presented in Sec. \ref{sec cal} is based
on the logarithmically dominant contribution of \( \nu \sim \log c \)
with \( c=\mathrm{max}(k_{B}T,\xi )/E_{F} \); it is a many-body effect,
related to the sharp edge of the Fermi surface, that occurs only in
the Cooper channel \( \mathbf{p}_{2}=-\mathbf{p}_{1} \). This approach
fails for in the situation where \( r_{s} \) is very small, when
the screening is too small to reduce the forward divergence of the
unscreened Coulomb potential. In this situation, one must consider
carefully the contribution of forward scattering intermediate states
with \( q=|\mathbf{k}-\mathbf{p}|\simeq 0 \) as they yield at large
term \( V\sim 2\pi e_{0}^{2}/k_{s} \). For such states, one must
keep the restriction \( k\simeq p\simeq k_{F} \), but consider the
contribution of small angles \( \phi =\angle (\mathbf{k},\mathbf{p}) \),
defined by \cite{smallanglebreak}\begin{equation}
\label{l}
\phi ,|\phi -\theta |\ll \bar{\phi }=\frac{d}{p}\to \frac{\delta }{p}=\frac{p-k_{F}}{p}\simeq \frac{c}{2}\ll 1,
\end{equation}
where \( d=k-k_{F} \), \( \delta =p-k_{F} \) and \( \theta =\angle (\mathbf{p},\mathbf{p}') \)
is the scattering angle. This corresponds to the forward scattering
into \emph{virtual} states; we do not yet specify the real scattering
angle \( \theta  \) between the initial and final states. We now
go back to the iterations of the Bethe-Salpeter, after the frequency
integration (\ref{log-div}), still considering the Cooper channel
only.

\subsection{Angular integral.}

We introduce the polar notation \( \mathbf{k}_{i}=(k_{i,}\phi _{i}) \)
and write the potential \( V(k_{1},\phi _{1};k_{2},\phi _{2})\equiv V(\mathbf{k}_{1}-\mathbf{k}_{2}) \),
assuming \( \mathbf{p}=(p,0),\mathbf{p}'=(p,\theta ) \). We first
consider the angular integrals for the n\( ^{\mathrm{th}} \) order
iteration of the Bethe-Salpeter equation:

\begin{widetext}\begin{equation}
\label{smallangul}
\frac{1}{2\pi }\int _{-\bar{\phi }}^{\bar{\phi }}d\phi _{1}V(p,\theta ;k_{1},\phi _{1})\frac{1}{2\pi }\int _{-\bar{\phi }}^{\bar{\phi }}d\phi _{2}V(k_{1},\phi _{1};k_{2},\phi _{2})\, \times ...\times \, \frac{1}{2\pi }\int _{-\bar{\phi }}^{\bar{\phi }}d\phi _{n-1}V(k_{n-2},\phi _{n-2};k_{n-1},\phi _{n-1})V(k_{n-1},\phi _{n-1};p,0)
\end{equation}
\begin{eqnarray*}
\simeq \left( \frac{\bar{\phi }}{\pi }\right) ^{n-1} & \underbrace{V(p,\theta ;k_{1},0)V(k_{1},0;k_{2},0)...} & \underbrace{V(k_{n-2},0;k_{n-1},0)V(k_{n-1},0;p,0)}\nopagebreak \\
 & \hspace {10mm}=a_{\theta }a^{n-3} & \hspace {25mm}=b(k_{n-2},k_{n-1})
\end{eqnarray*}
\end{widetext}

\subsection{Energy integral.}

First we consider \begin{equation}
\label{i}
b(k',k)=\frac{2\pi e_{0}^{2}}{k_{s}+|k'-k|}\frac{2\pi e_{0}^{2}}{k_{s}+|k-p|}.
\end{equation}
We truncate the energy integration to a small range \( k\in [k_{F}=p-\delta ,p+\delta ] \)
around \( p \) and define \( B(k'):=(1/2\pi )\int ^{k_{F}+2\delta }_{k_{F}}dkkD(k)b(k',k) \).
We can expand the lengthy result in lowest order in \( k_{s} \) and
find\begin{equation}
\label{i}
B(k')\simeq i\frac{m\pi ^{2}e_{0}^{4}}{\hbar ^{2}k_{s}|k'-p|}[1+\mathcal{O}(k_{s}/p)].
\end{equation}
 Thus, the subsequent \( k' \)--integration will be mainly given
by a small region around \( k'=p \), as expected. Hence one should
set \( k'=p \) in the exact result \( B(k') \) before expanding
in lowest order of \( k_{s} \), which yield\begin{equation}
\label{i}
B(k'=p)\simeq i\frac{m\pi ^{2}e_{0}^{4}}{\hbar ^{2}k_{s}^{2}}[1+\mathcal{O}(k_{s}/p)].
\end{equation}
For the integration of \( a(k) \) in \( A:=(1/2\pi )\int ^{k_{F}+2\delta }_{k_{F}}dkkD(k)a(k) \),
we set \( k'=p \) and expand the result; this gives\begin{equation}
\label{i}
A\simeq -i\frac{m\pi e_{0}^{2}}{2\hbar ^{2}k_{s}}[1+\mathcal{O}(k_{s}/p)].
\end{equation}
For the last term containing the scattering angle, \( A_{\theta }:=\frac{1}{2\pi }\int ^{k_{F}+2\delta }_{k_{F}}dkkD(k)a_{\theta }(k) \)
we have \begin{equation}
\label{i}
A_{\theta }\simeq =-i\frac{m\pi e_{0}^{2}}{2q\hbar ^{2}}[1+\mathcal{O}(k_{s}/p)]=A\frac{k_{s}}{q}
\end{equation}
with \( q=2k_{F}\sin |\theta /2| \). Writing the series \( t=\sum _{n}t^{(n)} \),
we find for the \( n^{\mathrm{th}} \)-order term (\( n>0 \)) 
\begin{equation}
\label{trs}
t^{(n)}=\left( \frac{\bar{\phi }}{\pi }\right)^{n-1} A^{n-1} A_\theta B =\frac{2\pi e_{0}^{2}}{q}D^{n-1},
\end{equation}
with
\begin{equation}
\label{i}
D= \frac{i}{8\pi} c=\frac{i}{8\pi}\frac{\xi}{E_F}.
\end{equation}

\subsection{Result}

So far we have only considered the case where all intermediate angles
are \( \phi _{i}=0 \), \( i=1,..n \), and only the final angle is
the scattering angle \( \theta  \). There are \( n \) equivalent
cases giving the same contribution, where the first \( j \) angles
are zero and the remaining ones are \( \phi _{i}=\theta ,j<i\le n \).
Thus the expansion series of the Bethe-Salpeter Eq. is \( t=V+\sum _{n>0}n\, t^{(n)} \).
Performing the summation, we get the scattering \( t- \)matrix for
Coulomb scattering in the limit of very small \( r_{s}\ll 1 \):\begin{equation}
\label{notforw}
t=\frac{2\pi e_{0}^{2}}{q+k_{s}}+\frac{2\pi e_{0}^{2}}{q}D\frac{2-D}{(1-D)^{2}}\stackrel{k_{s}\ll q}{=}\frac{2\pi e_{0}^{2}}{q}\frac{1}{(1-D)^{2}},
\end{equation}
\(  \)with \( q=2k_{F}\sin |\theta /2| \). We can further expand
this result in \( D\ll 1 \) \( \then \, t\simeq 2\pi e_{0}^{2}/q=V_{C} \);
however, this merely reflects the fact that the Born approximation
with unscreened potential is accurate in the limit \( r_{s}\to 0 \)
because \( e_{0}^{2} \) is also proportional to \( r_{s} \). We
also note that with this result the scattering amplitude has a phase
\( 2\mathrm{Arctan}|D|=2\mathrm{Arctan}(\bar{\phi }/8\pi ) \); however,
it is independent from the angle \( \theta  \) and therefore does
not yield quantum oscillations in the singlet/triplet scattering length
as discussed in Sec. \ref{quosc}.

In the case of forward scattering\cite{forwnote} with \( \theta =0 \),
the counting of equivalent arrangements of the intermediate states
gives a factor \( 2^{n} \) instead of \( n \). The result has a
more familiar form:\begin{equation}
\label{forw}
t=\frac{2\pi e_{0}^{2}}{k_{s}}\frac{1}{1-2D}.
\end{equation}
As expected, the forward scattering amplitude (\ref{forw}) is larger
than (\ref{notforw}) by a factor \( \sim q/k_{s}\sim 1/r_{s} \).

\subsection{Comparison}

We now compare the \( n^{\mathrm{th}} \) order obtained here, \( t^{(n)}_{r_{s}} \)
defined in Eq. (\ref{trs}) with the calculation of Sec. \ref{sec cal},
i.e. the \( m^{\mathrm{th}} \) Fourier coefficient \( t_{m}^{(n)}=v_{m}(\nu v_{m})^{n-1} \).
Their ratio is
\begin{equation}
\label{i}
\frac{t_{m}^{(n)}}{t_{r_{s}}^{(n)}}=\frac{v_{m}q}{n2\pi e_{0}^{2}}\left( \frac{\nu v_{m}}{D}\right) ^{n-1}\sim \frac{1}{n}\left[ \log \left( \frac{k_{s}}{k_{F}}\right) \right] ^{n}\left[ \frac{k_{s}}{k_{F}}\frac{\log c}{c}\right] ^{n-1}
\end{equation}
where have considered the small \( r_{s} \) approximation (\ref{vnap1})
\( v_{m}\simeq (2\pi e_{0}^{2}/k_{F})\log \left( k_{s}/k_{F}\right)  \).
Therefore, the calculation of Sec. \ref{sec cal} is valid provided
that \begin{equation}
\label{ii}
r_{s}|\log r_{s}|\gg \frac{\pi }{\sqrt{2}}\frac{c}{|\log c|}
\end{equation}
Taking \( c=\delta /k_{F}=0.01 \), we find that we need \( r_{s}>0.01 \),
which is always the case for typical semiconductor material. In conclusion,
the large value of \( r_{s} \) does not allow one to see the contribution
of forward scattering into  (virtual or real) states.


\begin{thebibliography}{10}
\bibitem{Top00}M. A. Topinka \emph{et al}., Science \textbf{289}, 2323 (2000); Nature
(London) \textbf{410}, 183 (2001).
\bibitem{Roy02} B.J. Leroy, J. Phys.:Condens. Matter \textbf{15}, R1835 (2003).
\bibitem{Sar04} D.S. Saraga, B.L. Altshuler, D. Loss, and R.M. Westervelt, Phys. Rev. Lett. {\bf 92}, 246803 (2004).
\bibitem{FW71}A.L. Fetter and J.D. Walecka, \emph{Quantum Theory of Many-Particle
Systems} (McGraw-Hill, 1971).
\bibitem{Mah00}G. D. Mahan, Many-particle physics (Plenum, New York, 2000).
\bibitem{Ste67}F. Stern, Phys. Rev. Lett. \textbf{18}, 546 (1967).
\bibitem{Giu82}G.F Giuliani and J.J. Quinn, Phys. Rev. B \textbf{26}, 4421 (1982).
\bibitem{Zhe96}L. Zheng and S. Das Sarma, Phys. Rev. B \textbf{53}, 9964 (1996).
\bibitem{Bur00}G. Burkard, D. Loss, and E.V. Sukhorukov, Phys. Rev. B \textbf{61},
R16303 (2000).
\bibitem{Gal04}V.M. Galitski and S. Das Sarma, Phys. Rev. B {\bf 70}, 035111 (2004)
\bibitem{Tin75}C.S. Ting, T.K. Lee, and J.J. Quinn, Phys. Rev. Lett. \textbf{34},
870 (1975).
\bibitem{Lut66}W. Kohn and J.M Luttinger, Phys. Rev. Lett. \textbf{15}, 524 (1966);
J.M. Luttinger, Phys. Rev. \textbf{150}, 202 (1966).
\bibitem{Chu93}A. V. Chubukov, Phys. Rev. B \textbf{48}, 1097 (1993).
\bibitem{Suw04}T. Suwa, K. Takayanagi, and E. Lipparini, Phys. Rev. B \textbf{69},
115105 (2004). In contrast to our work, these authors consider a finite
total momentum \( P=k_{F}\sqrt{2}\neq 0 \) (for particles within
the Fermi sea), a 3D-like screening \( V(r)\sim e^{-\mu r}/r \),
large values of \( r_{s}\sim 1-20 \), and use a Bethe-Goldstone equation
(which does not include the exchange process shown in our Fig. \ref{manypart}).
\bibitem{Bon98}M. C. Bønsager, K. Flensberg, B. Y.-K. Hu, and A. H. MacDonald, Phys.
Rev. B \textbf{57}, 7085 (1998).
\bibitem{Jal89}R. Jalabert and S. Das Sarma, Phys. Rev. B \textbf{40}, 9723 (1989).
\bibitem{Gal03}V. M. Galitski and S. Das Sarma, Phys. Rev. B \textbf{67}, 144520
(2003).
\bibitem{Los98}D. Loss and D.P. DiVincenzo, Phys. Rev. A \textbf{57}, 120 (1998).
\bibitem{BurLos} G. Burkard and D. Loss, in {\em Semiconductor Spintronics and Quantum Computation} (Springer-Verlag, New York, 2002).
\bibitem{Van03}
L.M.K. Vandersypen {\em et al.}, in {\em Quantum Computing and Quantum Bits in Mesoscopic Systems} (Kluwer Academic/Plenum, New York, 2003).
\bibitem{Rec01}P. Recher, E.V. Sukhorukov and D. Loss, Phys. Rev. B \textbf{63},
165314 (2001).
\bibitem{Oli02}W.D. Oliver, F. Yamaguchi, Y. Yamamoto, Phys. Rev. Lett. \textbf{88},
037901 (2002).
\bibitem{Sar03}D.S. Saraga and D. Loss, Phys. Rev. Lett. \textbf{90}, 166803 (2003).
\bibitem{Rec02}P. Recher and D. Loss, Phys. Rev. B \textbf{65}, 165327 (2002); C.
Bena, S. Vishveshwara, L. Balents, M.P.A. Fisher,Phys. Rev. Lett.
\textbf{89}, 037901 (2002).
\bibitem{Les01}G.B. Lesovik, T. Martin and G. Blatter, Eur. Phys. J. B \textbf{24},
287 (2001); G. Falci, D. Feinberg, and F.W.J. Hekking, Europhys. Lett.
\textbf{54}, 255 (2001).
\bibitem{Rec03}P. Recher and D. Loss Phys. Rev. Lett. \textbf{91}, 267003 (2003)
\bibitem{Cos01}A.T. Costa, Jr. and S. Bose, Phys. Rev. Lett. \textbf{87}, 277901
(2001); S. Bose and D. Home, Phys. Rev. Lett. \textbf{88}, 050401
(2002).
\bibitem{Sam03}P. Samuelsson, E.V. Sukhorukov, and M. B\"uttiker, Phys. Rev. Lett.
\textbf{91}, 157002 (2003).
\bibitem{Bee03}C.W.J. Beenakker, C. Emary, M. Kindermann, and J.L. van Velsen, Phys.
Rev. Lett. \textbf{91}, 147901 (2003).
\bibitem{Sam04}P. Samuelsson, E. V. Sukhorukov, and M. Büttiker, Phys. Rev. Lett.
\textbf{92}, 026805 (2004).
\bibitem{Tay72}J.R. Taylor, \emph{Scattering Theory} (John Wiley \& Sons, 1972).
\bibitem{Jac01}J. Jacoby, Phys. Scripta \textbf{64}, 220 (2001).
\bibitem{Anderson}From this perspective, the experimental demonstration of the scattering
of quasi-particle states would also shed light on the
applicability of Fermi liquid theory in 2D; see P.W Anderson, cond-mat/0101417;
Phys. Rev. Lett. \textbf{71}, 1220 (1993) and reference {[}1{]} therein.
\bibitem{Note momenta}For simplicity we call the wave-vectors (\( \mathbf{p},\mathbf{k},\mathbf{q},.. \))
'momenta', and set \( \hbar =1 \) for the frequencies \( \omega  \).
\bibitem{Bar83}G. Barton, Am. J. Phys. \textbf{51}, 420 (1983); S.K. Adhikari, \emph{ibid}.
\textbf{54}, 362 (1986).
\bibitem{3dlines}It is only the motion of the electron that is confined in 2D; the
field lines are 3D.
\bibitem{And82}T. Ando, A.B. Fowler, and F. Stern\emph{,} Rev. Mod. Phys \textbf{54},
437 (1982); C.W.J. Beenakker and H. van Houten, Solid State Physics
\textbf{44}, 1 (1991).
\bibitem{renorma}One can easily consider \( Z \) and \( m^{*} \) by rescaling the
quantity \( \nu  \) defined in Eq. (\ref{log-div}) via \( \nu \to \nu \times Z^{2}m^{*}/m. \)
\bibitem{Chu89}A.V. Chubukov and M. Yu Kagan, J. Phys.: Condens. Matter \textbf{1},
3135 (1989).
\bibitem{linear}More formally, one must consider a linear expansion of the denominator
of \( V \) to make the integral convergent for large \( k \), as
the decay of \( D(k) \) is too slow. Alternatively, one can put a
cut-off for the integral at, e.g., \( 2k_{F} \).
\bibitem{Naz03}See e.g. \emph{Quantum Noise in Mesoscopic Physics}, NATO ASI Series
II, Vol. 97, ed. Yu. V. Nazarov (Kluwer, 2003).
\bibitem{Les94}G.B. Lesovik and L.S. Levitov, Phys. Rev. Lett. \textbf{72}, 538 (1994).
\bibitem{Bur03}J.C. Egues, G. Burkard, and D. Loss, Phys. Rev. Lett. \textbf{89},
176401 (2002); G. Burkard and D. Loss, Phys. Rev. Lett. \textbf{91},
087903 (2003). 
\bibitem{Kaw01}S. Kawabata, J. Phys. Soc. Jpn. \textbf{70}, 1201 (2001).
\bibitem{Gran04}P. Recher, D.S. Saraga, and D. Loss
in {\em Fundamental Problems of Mesoscopic Physics: Interactions and Decoherence}, NATO Science Ser. II, Vol. 154 (Kluwer, Dordrecht, 2004). 
\bibitem{Han03} P. Recher, E. V. Sukhorukov,
D. Loss, Phys. Rev. Lett. \textbf{85}, 1962 (2000);
R. Hanson, L.M.K. Vandersypen, L.H. Willems van Beveren, J.M. Elzerman,
I.T. Vink, L.P. Kouwenhoven, cond-mat/0311414.
\bibitem{Pot02}R.M. Potok \emph{et al.,} Phys. Rev. Lett. \textbf{89}, 266602 (2002). 
\bibitem{Fie99}R. Fiederling \emph{et al}., Nature \textbf{402}, 787 (1999); Ohno
\emph{et al}., Nature \textbf{402}, 790 (1999).
\bibitem{Nor99} A. North {\em et al.}, IEEE J. Quantum Elect. {\bf 35}, 352 
(1999).
\bibitem{Tar00}S. Tarucha \emph{et al.}, Phys. Rev. Lett \textbf{84}, 2485 (2000).
\bibitem{ioffe}V.F. Gantmakher and Y.B. Levinson, \emph{Carrier Scattering in Metals
and Semiconductors} (North-Holland, Amsterdam, 1987); See also http://www.ioffe.rssi.ru/ SVA/NSM/Semicond/GaAs/basic.html. 
\bibitem{Pri81}P.J. Price, Ann. Phys. \textbf{133}, 217 (1981).
\bibitem{oscill-negl}This is similar to neglecting the oscillating term in the Poisson
sum \( \delta (\theta =\pi )=\sum _{n}(-1)^{n}= 0 \).
\bibitem{smallanglebreak}The calculation in Sec.\ref{sec cal} involved the expansion 
\( |\mathbf{k}-\mathbf{p}|=
\sqrt{p^{2}+k^{2}-2pk\cos \phi }\simeq 
2p\sin (\phi /2)
\left\{ 1+d/p+ O\left( d^2/\left[2 p \sin(\phi /2)\right]^2 \right)\right\}  \),
which breaks down for small angles \( \phi  \).
\bibitem{forwnote}We note that the present calculation is not strictly valid in the
case of forward scattering, because of the remaining terms 
\( O\left( d^{2}/[2 p \sin(\phi /2)]^2\right)  \)
in (\ref{smallangul}) which only give a negligeable contribution
in the integration when \( \bar{\phi }\ll 2\tan \theta /2\simeq \theta  \)
(preventing the limit \( \theta \to 0 \)). \end{thebibliography}
\end{document}